\documentclass[bibyear]{aa}
\usepackage{graphicx}
\usepackage{lipsum}
\usepackage{float}
\usepackage{natbib}
\usepackage{amsmath}
\usepackage{amssymb}
\usepackage{multicol}
\usepackage{stfloats}
\usepackage{color}
\usepackage{textcomp}
\usepackage[flushleft]{threeparttable}
\usepackage[dvipsnames]{xcolor}
\usepackage{txfonts}
\usepackage{yfonts}
\usepackage[T1]{fontenc}
\usepackage[utf8]{inputenc}
\usepackage{url}
\usepackage{hyperref}

\hypersetup{
    colorlinks=true,
    citecolor = blue,
    linkcolor=blue
            }

\usepackage{xcolor}

\usepackage{subfigure}

\newcommand{\degree}{$^\circ$\,}

\usepackage[T1]{fontenc}
\usepackage{pdfcomment}
\usepackage[normalem]{ulem}

\begin{document}

    \title{
TESS flares and chromospheric variability in the Pleiades, the Hyades, and Praesepe 
}
   \author{A.~Görgei
          \inst{1,2,3}
          \and
          K.~Vida
          \inst{1,2}
          \and
          B.~Seli
          \inst{1,2}
          \and
          K.~Oláh
          \inst{1,2}
          \and
          P.~Sági
          \inst{1,2}
          \and
          B.~Schmercz 
          \inst{1,2,3}
          \and
          R.~M.~Roettenbacher
          \inst{4}
          \and
          Zs.~K\H{o}v\'ari
          \inst{1,2}
          }

   \institute{Konkoly Observatory, HUN-REN Research Centre for Astronomy and Earth Sciences, Konkoly Thege \'ut 15-17., H-1121 Budapest, Hungary\\
              \email{gorgei.anna@csfk.org}
        \and
            CSFK, MTA Centre of Excellence, Konkoly Thege \'ut 15-17., H-1121 Budapest, Hungary
         \and
             E\"otv\"os Lor\'and University, Institute of Physics and Astronomy, Department of Astronomy, P\'azm\'any P\'eter s\'et\'any 1/A, H-1117 Budapest, Hungary
        \and
        Department of Astronomy, University of Michigan, Ann Arbor, MI 48109, USA
             }

   \date{Received ...; accepted ...}

  \abstract
 {Magnetic activity is a result of the interaction between rotation and dynamo processes in cool stars with outer convection zones. Since open cluster members share a common age, distance, and chemical composition, they provide an opportunity to study the magnetic activity of cool stars in a controlled astrophysical context.}
 {We used 120-s cadence photometric observations from the Transiting Exoplanet Survey Satellite (TESS) to study the flare activity of stars in three open clusters with different ages:  the Pleiades (100\,Myr), the Hyades (700\,Myr) and Praesepe (700\,Myr). We used the Ohio State Multi-Object Spectrograph (OSMOS) at MDM Observatory with the goal of identifying coronal mass ejections (CMEs) on stars in the Pleiades and Praesepe.}
 {We investigated the flaring characteristics of the cluster members using the TESS light curves: we identified the flares, then divided the dataset into different categories based on the effective temperature of the flaring stars and the amplitude of their light curves after flare removal. We calculated the flare energies and investigated the flare frequency distributions of each previously defined group. We studied the dynamic spectra of the H$\alpha$ line and its environment for the stars of the Pleiades and Praesepe to search for indication of CMEs. In order to quantify temporal variations in the line strength we measured the H$\alpha$ chromospheric activity indices for each spectrum.}
 {The power-law indices of the flare frequency distributions obtained in our analysis span the 1.8$-$2.8 range. 
 We found the highest flare rates in the Pleiades for all investigated spectral types. Our results show that the flare rate of M dwarfs is similar in the Pleiades and the Hyades while for K-type stars the flare rate in the Pleiades is an order of magnitude higher. 
 Our analysis reveals that the flare rate in M stars changes with their effective temperature: the rate is highest for $T_{\rm eff}\sim$3200$-$3300\,K
 ---coinciding with the transition between partially and fully convective stars--- with smooth decrease in flare rate towards cooler and hotter M dwarfs. We detected no CME signals in the MDM/OSMOS spectra, and we used this information to place an upper limit on the CME rate.}
 {}


   \maketitle

\section{Introduction}

Magnetic activity is a fundamental property of cool stars with outer convection zones, driven by the interplay between rotation and dynamo processes \citep[e.g.,][]{2017LRSP...14....4B,2023SSRv..219...70I}.
Partially convective stars possess a radiative core and convective envelope, with magnetic fields thought to be generated primarily by an $\alpha\Omega$ dynamo operating in the tachocline, where strong rotational shear exists between the two regions \citep[e.g.,][]{2023SSRv..219...35C}. In contrast, fully convective stars lack a radiative core and therefore do not possess a tachocline, with their magnetic fields generated by a distributed turbulent ($\alpha^2$) dynamo operating throughout the convection zone \citep{1980mfmd.book.....K}. Both partially and fully convective stars can exhibit strong magnetic activity, including starspots, flares, and coronal mass ejections. Observations show that magnetic activity remains strong across the fully convective boundary \citep[e.g.,][]{2006A&A...460L..35S,2011ApJ...731...63G,2016ApJ...828L..22S,2023ApJ...954..142L,2024A&A...687A.138I}, suggesting that stellar dynamos can operate efficiently despite significant changes in internal structure.
In young stellar populations, rapid rotation enhances magnetic field generation, leading to elevated levels of activity \citep[e.g.,][]{2004A&ARv..12...71G,2007ApJ...662.1254S,2010A&A...520A..15M}. The magnetic phenomena not only trace the underlying dynamo but also play a key role in shaping stellar angular momentum evolution \citep[e.g.,][]{1972ApJ...171..565S, 1988ApJ...333..236K, 2015ApJ...799L..23M} and influencing the circumstellar environment \citep[e.g.,][]{2016A&A...594A..95A,2020IJAsB..19..136A, 2021LRSP...18....3V}. Flares represent the most readily observable signatures of magnetic activity. Their occurrence rates and energy distributions encode information about the magnetic energy budget and its release processes \citep[e.g.,][]{2004A&ARv..12...71G,2006SoPh..234...41K,2011soca.book.....A,2015ApJ...809...79O,2015RSPTA.37340259T}. On the Sun large flares are also often associated with coronal mass ejections (CMEs) \citep[e.g.,][]{2005JGRA..11012S05Y,2007JGRA..112.6112G,2012LRSP....9....3W}, however, direct detections of stellar CMEs remain challenging \citep{2020MNRAS.493.4570L, 2024Univ...10..313V}. Stellar CME detection techniques include the search for Doppler-shifted Balmer line emission during flares \citep[e.g.,][]{2019A&A...623A..49V}, X-ray/EUV dimming after flares \citep[e.g.,][]{2021NatAs...5..697V}, and radio bursts \citep[e.g.,][]{2025Natur.647..603C}. These methods are limited by the trade-off between signal-to-noise ratio (S/N), spectral resolution and integration time, as the CME signal is usually weak and changes rapidly. \cite{2020MNRAS.494.3766O} modeled how detectable stellar CMEs are through Balmer-line spectroscopy. They found that mid–late M dwarfs and highly active stars are the best targets, but CME detections remain rare because of observational biases and unfavorable geometry.


Open clusters provide an ideal laboratory for studying the magnetic activity of cool stars in a controlled astrophysical context. Their members share a common age, distance, and chemical composition, allowing the study of activity trends as a function of stellar parameters. Of particular interest due to their proximity to Earth are Praesepe \citep[M44, e.g.,][]{2009A&A...497..209V,2023A&A...677A.163A}, the Hyades \citep[Collinder\,50, Melotte\,25, e.g.,][]{1998A&A...331...81P,2009A&A...497..209V,2011AJ....141..172M}, and the Pleiades \citep[M45, e.g.,][]{2009A&A...497..209V,2023A&A...677A.163A}. These clusters fall within a key age range between 100\,Myr and 1\,Gyr, during which the rotation and magnetic activity of stars undergo significant evolution \citep{1997ApJ...483..947G, 2005ApJS..160..390P, 2013A&A...556A..36G, 2015A&A...577A..98G}. The stars leave the saturated regime, where activity indicators remain nearly constant despite decreasing rotation rate, to the unsaturated \cite[or interface-sequence,][]{2003ApJ...586..464B} regime, in which magnetic activity declines progressively as stellar rotation slows down \citep{2010ARA&A..48..581S, 2011ApJ...743...48W, 2012LRSP....9....1R}.

The advent of space-based photometric missions such as the Transiting Exoplanet Survey Satellite \citep[TESS,][]{tess} has revolutionized the study of stellar activity by providing high-cadence, long-duration light curves for a large sample of stars. These data enable systematic investigations of flare occurrence and energetics over a wide range of stellar ages and activity levels. At the same time, ground-based multi-object spectroscopy allows simultaneous monitoring of multiple cluster members, offering a unique opportunity to search for spectroscopic signatures of CMEs, such as transient asymmetries in chromospheric emission lines \citep[e.g., in the H$\alpha$ and H$\beta$ profiles,][]{2025ApJ...991...54C}.

In this work, we combined TESS photometry with multi-object spectroscopic observations to investigate magnetic activity of members of the three nearby open clusters mentioned above---the Pleiades, the Hyades and Praesepe (with early results presented in the conference proceeding of \citealt{2024csss.confE..63V}). We focused on flare properties derived from high-cadence light curves and complemented these with spectroscopic diagnostics. By targeting carefully selected cluster members with high flare rates, we aimed to maximize the likelihood of detecting CME signatures and to place observational constraints on their occurrence. The paper is organized as follows: in Sect.\,\ref{sect:obs}, we describe the photometric and spectroscopic observations, including target selection. Sect.\,\ref{sect:photometry} presents the photometric analysis, including flare detection and energy determination, as well as flare frequency distributions. In Sect.\,\ref{sect:spectroscopy}, we analyze spectroscopic variability with a focus on H$\alpha$ line profiles. The results are discussed in Sect.\,\ref{sect:discussion} and our conclusions are summarized in Sect.\,\ref{sect:conclusions}.

\section{Observations}\label{sect:obs}

\subsection{Photometric observations} \label{section:phot_obs}

To study the flare activity of open cluster members, we used 120\,s cadence photometric observations from TESS. These are suitable for investigating short-term brightness changes on timescales ranging from a few minutes to several days. (e.g., effects of flares and spots). TESS has been in a 13.7-day-long orbit around Earth since its launch in April 2018. The instruments on board are four 10.5\,cm telescopes that cover 24\degree $\times$ 96\degree instantaneous strips (sectors). TESS operates in nearly continuous sectors, moving every $\sim$27\,d and has covered nearly all of the sky since its launch. The data acquisition is interrupted only by data transfer to Earth causing short gaps in the light curves.

Members of the Pleiades, the Hyades and Praesepe clusters were observed in 13, 48, and 13 sectors, respectively. The cluster membership was taken from the publication of \cite{2025A&A...694A.258R} who have studied membership and tidal trails in more than a hundred open clusters.
With the aim of flare identification and analysis, we used the 120\,s cadence pre-search data conditioning simple aperture photometry (\mbox{PDCSAP}) light curves provided by the Science Processing Operations Center \citep[SPOC;][]{10.1117/12.2233418}.

In the case of the stars selected for spectroscopic observations, we used 20\,s cadence TESS light curves observed as part of the guest observer program G06138 (PI: R. Roettenbacher). This cadence is enough to track the detailed temporal morphology of flares to aid the spectroscopic observations, and to detect short events.

\subsection{Spectroscopic observations} 
The goal of the spectroscopic observations was to identify coronal mass ejections on stars with known ages. For this purpose, our best option was to pick magnetically active stars that are most likely to exhibit (super)flares and CMEs from young open clusters. The most suitable ones were the nearby clusters as we needed to achieve high S/N and good time resolution simultaneously. The possible clusters were the Hyades, Praesepe, and the Pleiades. As the Hyades is the least dense out of the three, we choose Praesepe and the Pleiades to maximize the number of stars in the field-of-view (FoV) of the multi-object spectrograph. The member stars were determined based on the catalog of \cite{2021A&A...645A..42I}.

After selecting the two young, nearby open clusters Pleiades and Praesepe, with ages of 100 and 700\,Myr \citep{2016A&A...585A.150N}, we aimed to maximize our chances of
detecting a CME by observing as many simultaneous targets as possible. For this purpose, we first determined the flare rates of member stars from light curves obtained by the Kepler/K2 mission \citep{2014PASP..126..398H} -- the extended mission of the Kepler space telescope \citep{2010ApJ...713L..79K} -- and TESS and selected flaring stars brighter than $T_{\textrm{TESS}}=15^m$ (TESS magnitude) using the preliminary version of the flare catalog of \cite{Seli2025A&A...694A.161S}. After this step, we carried out a grid search in RA–Dec to find the optimal pointing for a FoV of 18’ diameter that maximized the chances of flare occurrence. The best possibility included 10 stars from Praesepe and 12 from the Pleiades in their respective FoV.

We used the Ohio State Multi-Object Spectrograph (\mbox{OSMOS}) on the $2.4$\,m Hiltner telescope at MDM Observatory (Arizona, US) with the moderate resolution (R) volume phase holographic (VPH) grism (R $\approx$ 1600) for 10 days in 2023 November. The observations covered the $600-900$\,nm wavelength range.
Masks were designed using version 1.6  of the OMS (OSMOS Mask Simulator) mask design software package\footnote{ \url{https://www.astronomy.ohio-state.edu/MDM/OSMOS/oms.html} } installed in a Docker environment. 
Typical CMEs have velocities ranging from a few hundred to a few thousand\,km\,s$^{-1}$, thus, the setup was determined in a way to have a velocity resolution ($\approx$180\,km\,s$^{-1}$) that can resolve larger CMEs with good S/N while also having acceptable temporal resolution. As on most nights the weather did not allow for observations, for the Pleiades and Praesepe open clusters we had 12 and 15 hours on-target time, respectively. To allow for time resolution of possible CMEs the exposure time was 15 minutes. The timing of the observation allowed for co-temporal light curves obtained by TESS as part of the guest observer program G06138.

The observed spectra were reduced with the standard reduction steps, including dark subtraction, tracing and extraction of spectra, sky background removal and wavelength calibration with a HgNe spectral lamp.

\section{Photometric analysis}\label{sect:photometry}

We investigated the flaring characteristics of the cluster members using the PDCSAP light curves in  the available TESS sectors.  When preparing for the flare identification we compared a hand selected flare sample with a machine learning based sample of \cite{2018A&A...616A.163V} made with \texttt{flatwrm}. For problematic light curves with instrumental artifacts, and in the case of low-energy flares we could get more accurate flare numbers when searching by eye. As the completeness was important and the number of light curves was manageable, the flares were identified by visual inspection of the light curves. We note that as TESS continues to operate the noise level of the light curves change, especially for later sectors.  This hinders the detection of smaller flares, but since the flare-finding method is not directly tied to the photometric scatter, we estimate that its effect on the flare statistics is negligible.

\begin{figure}
    \centering
    \includegraphics[width=1\linewidth]{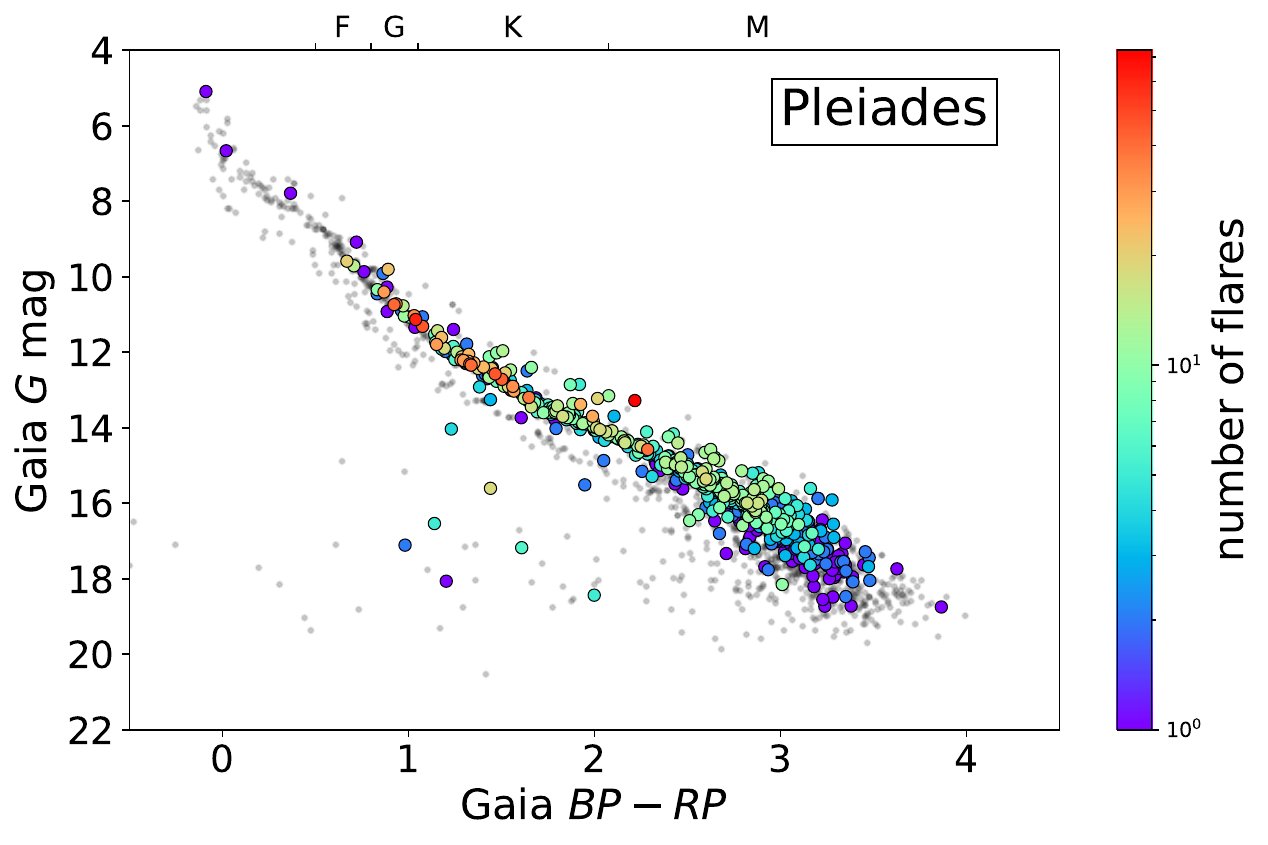}
    \includegraphics[width=1\linewidth]{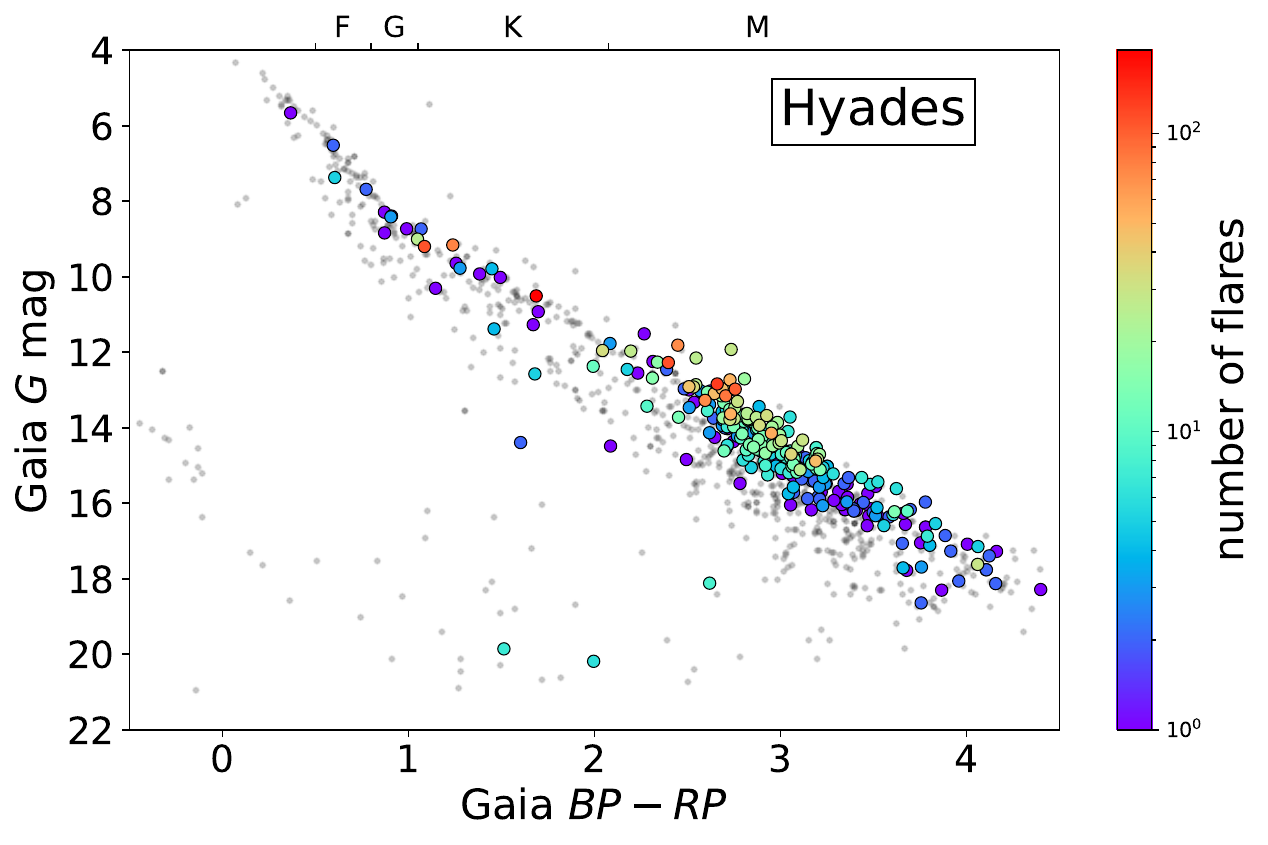}
    \includegraphics[width=1\linewidth]{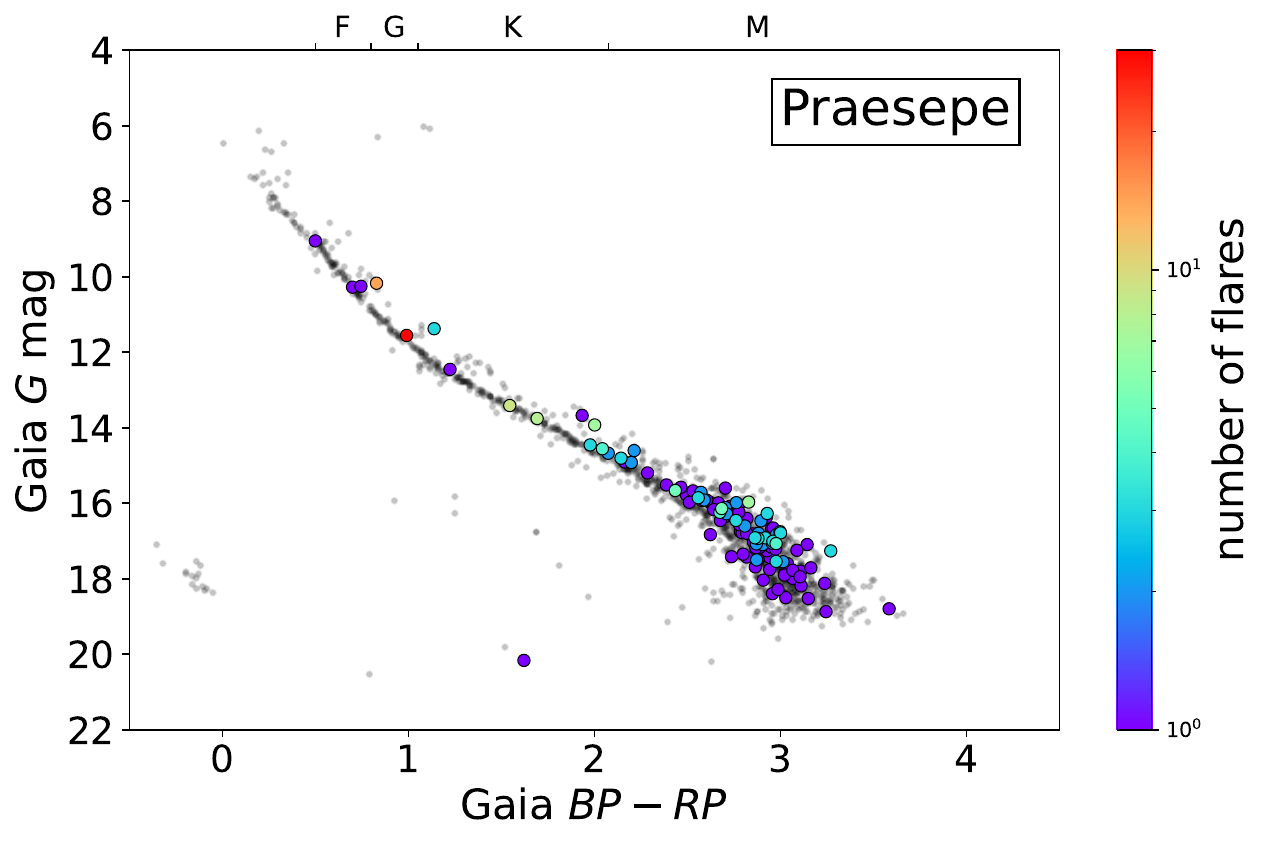}
    \caption{The HRD of the three open clusters with extinction corrected Gaia $BP-RP$ index and Gaia $G$ magnitude. The three panel from the top to the bottom are corresponding to the Pleiades, the Hyades and Praesepe. All cluster members are marked with small gray dots, while the flaring ones are colored according to their flare number.}
    \label{fig:hrds}
\end{figure}

The total number of flares and the rate of the flaring stars is shown in Table \ref{tab:lc_and_flares} while their distribution on the Hertzsprung--Russell diagram (HRD) is presented in Fig.\ref{fig:hrds}. Most of the flaring members are late-type stars, and the equal mass binary sequence is also prominent above the main sequence, especially in the Pleiades.

After identifying the flares, we divided the dataset in two different ways: a) we defined multiple spectral groups based on the effective temperature of the flaring stars -- that were obtained from the TESS Input Catalog v8.2 \citep[TIC v8.2,][]{2019AJ....158..138S}, where available, or interpolated based on the Gaia BP$-$RP index --, and b) we split the stars into two groups based on the amplitude of their light curves. The color excess ($E(B-V)$) values that we used in the extinction correction were adopted from \cite{2018A&A...616A..10G} and are $0.045$, $0.001$ and $0.027$ for the Pleiades, Hyades and Praesepe, respectively. Table \ref{tab:flares_types} contains the number of flares for every cluster in each category and it also shows the temperature ranges assigned to the spectral types. The amplitudes of the light curves were calculated after masking the flares occurring on them. They were determined as the 95th percentile of the light curve values normalized with their median. This results in a lognormal distribution of amplitudes, the threshold was the median of this distribution for each cluster separately (0.056 for Pleiades, 0.013 for Hyades, and 0.056 for Praesepe). Table \ref{tab:amp} contains the detailed numbers on how many of the low and high amplitude stars belong to each stellar type. As shown in the two tables above, we found an order of magnitude fewer flares in Praesepe than in the other two clusters. This cluster is the most distant of the three analyzed, which makes its center appear denser on the photometric images than that of the others. Light curves from this region need more complex processing to be free of all contamination \citep{2026AJ....171..163S}, so \mbox{PDCSAP} light curves are not available. Because of the complexity of custom light-curve extractions, it is out of the scope for our project.
We note that F and G spectral groups for Praesepe do not contain enough flares for statistical analysis, as only two flares belong to the F group, and the G types are dominated by flares from one star. 

\begin{table*}[]
\caption{The table contains for each cluster their age, distance, metallicity, the total observation time of their stars, the number of stars with flares in them, the number of flares on their members, the daily flare rate and the rate of flaring stars among their members. Ages and metallicity values were adopted from \cite{2016A&A...585A.150N}, distances for the Pleiades and Praesepe from \cite{2018A&A...618A..93C} and for the Hyades from \cite{1998A&A...331...81P}.}
\label{tab:lc_and_flares}
\begin{center}
\begin{tabular}{ccccccccc}
\hline\hline
Cluster &Age  &Distance &Metallicity& Obs. time  & \textnumero & \textnumero & Flares & Rate of  \\\smallskip
 & [Myr] & [pc] & [Fe/H]& [d] &flaring stars & flares & [d$^{-1}$]&flaring stars \\ \hline
Pleiades & 100 & 135.6 & $-0.01\pm$0.05  & 169344 & 835   & 4573 &  0.027 & 0.558 \\
Hyades   & 720 & 46.3  &  0.13$\pm$0.02 & 75708  & 317   & 3399  & 0.045 & 0.475 \\
Praesepe & 730 & 185.5 &  0.16$\pm$0.00 & 80028  & 125   & 251   & 0.003 & 0.133 \\ \hline
\end{tabular}
\end{center}
\end{table*}

\begin{table*}[]
\caption{The table contains for each cluster the number of stars, flaring stars and identified flares in F, G, K, M spectral types and in the two amplitude categories.}
\label{tab:flares_types}
\begin{center}
\begin{tabular}{cc|ccc|ccc|ccc}
\hline\hline
         & $T_\text{eff}$ & \multicolumn{3}{c|}{Pleiades} & \multicolumn{3}{c|}{Hyades} & \multicolumn{3}{c}{Praesepe} \\ 
         & [K]  &\textnumero & \textnumero{} flaring& \textnumero & \textnumero  & \textnumero{} flaring& \textnumero & \textnumero & \textnumero{} flaring& \textnumero \\ \smallskip
         &   & stars & stars& flares & stars & stars&  flares & stars & stars& flares \\          \hline
F        & 6000--7500 & 75     & 5    & 47    & 50  & 3   & 8     & 78   & 2   &   2    \\ 
G        & 5200--6000 & 79     & 21   & 233   & 39  & 6   & 9     & 56   & 3   &   45   \\
K        & 3700--5200 & 218    & 159  & 1559  & 116 & 20  & 484   & 182  & 9   &   37   \\
M        & 2800--3700 & 1005   & 619  & 2734  & 409 & 275 & 2898  & 577  & 107 &   167  \\
high amp & ---        & 721    & 349  & 1022  & 319 & 216 & 2255  & 459  & 83  & 124  \\
low amp  & ---        & 722    & 455  & 3553  & 320 & 88  & 1141  & 459  & 38  & 127   \\ \hline
\end{tabular}
\end{center}
\end{table*}

\begin{table}[]
\caption{Table showing the number of different stellar types in the low and high light curve amplitude categories. The amplitudes are calculated after flare removal.}
\label{tab:amp}
\centering
\begin{tabular}{c|cc|cc|cc}
\hline\hline
  & \multicolumn{2}{c|}{Pleiades} & \multicolumn{2}{c|}{Hyades} & \multicolumn{2}{c}{Praesepe} \\
  & low          & high          & low         & high         & low          & high          \\ \hline
F &  5           &  0            &  3          &  0           &  2           &  0             \\
G & 20           &  1            &  6          &  0           & 3            &  0             \\
K & 155          &  4            &  11         &  9           & 8            &  1             \\
M & 275          &  344          &  68         &  207         &  25          &  82           \\ 
\hline
\end{tabular}
\end{table}

\subsection{Flare energies} \label{sect:flares}

When calculating the flare energies, we used the same approach as in \cite{Seli2025A&A...694A.161S} and \cite{2026A&A...706A..54G}. The initial step of the method is marking the flaring data points and normalizing the light curves, which is followed by fitting a  fourth-order polynomial to the light-curve segments bordering each flare for 30 minutes on either side. 
Next, the fourth-order polynomial baseline was removed, and we calculated the area below each flare (equivalent duration; ED) by integrating according to Simpson's rule.

The flare energies ($E_\mathrm{TESS}$) in the TESS band are calculated using the following relation:
\begin{equation}
\hspace{0.35\linewidth}
E_{\text{TESS}} = L_{\text{TESS}} \times ED,\quad 
\end{equation}
where $L_{\text{TESS}}$ is the luminosity of the target in the TESS band. To determine  $L_{\text{TESS}}$ for a given star, we interpolated a  BT-NextGen spectrum \citep{nextgen} grid to get a model spectrum corresponding to its effective temperature ($T_{\text{eff}}$), and surface gravity ($\log g$). Then we integrated the spectra over the entire wavelength range with and without multiplying with the TESS response function to calculate the ratio of TESS to bolometric luminosity ($L_{\text{bol}}$). To get the value of $L_{\text{bol}}$, we used the method described by \cite{gaia_bolcorr} and data from \cite{gaia_dr3} and applied the bolometric correction on the absolute magnitude of the star.

\subsection{Flare frequency distribution}

After calculating the energies of all flares, we investigated the flare frequency distributions (FFD) of each previously defined group. FFDs  are widely used to illustrate how often flares with different energies occur. From the slope of the trend they show, we can infer whether the energy release is dominated by small or large flares. At low energies, flare events have a lower recovery rate due to the noise of the light curves; therefore, only flares above a certain energy were included when analyzing the the FFD slopes. This energy threshold  was different for every group, since the apparent brightness of the stars depend both on their spectral type and their distance, its value falls within the log$E$ range of 32.8--34.5. Above this threshold, the FFDs were fitted using a linear function based on the following equation:
\begin{equation}
\hspace{0.3\linewidth} 
\log \nu(E) = (1 - \alpha_{\textrm{pl}}) \log{E} + b,
\end{equation}
where $\nu(E)$ is the cumulative number of flares with energy greater than $E$ , $\alpha_{\textrm{pl}}$ is the power law index and $b$ is the normalizing factor for the power law. We note, that when calculating flare rates, the durations of light curves without flares were also included in the reference time interval. We note, that the incomplete recovery of low-energy flares does not substantially affect the derived FFD slopes, because the power-law fits were performed only above the adopted completeness threshold, where the flare detection efficiency is expected to be unity. The incompleteness primarily affects the normalization of the FFD and the low-energy turnover, rather than the slope.

\begin{figure*}
    \centering
    \includegraphics[width=0.95\linewidth]{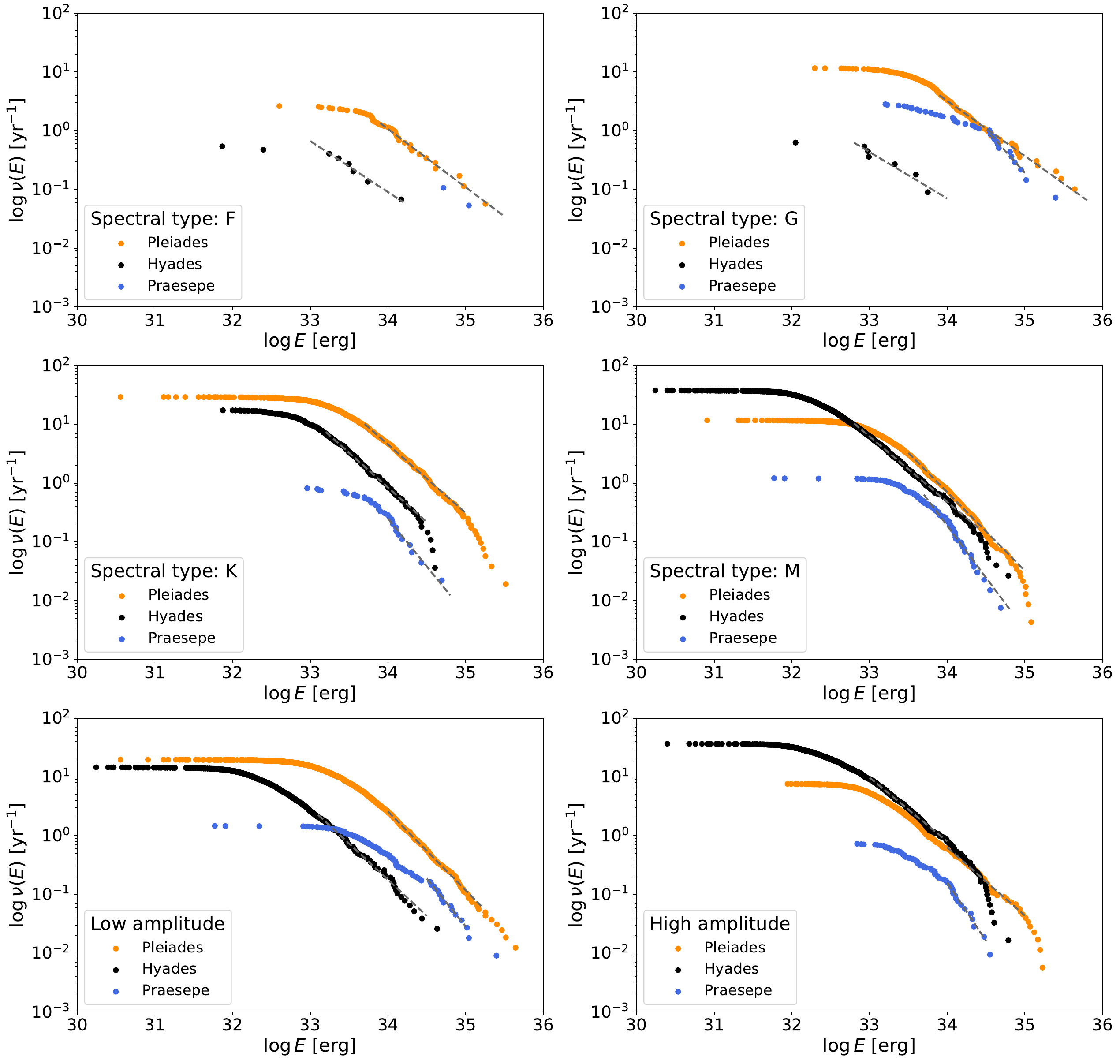}
    \caption{Flare frequency distribution diagrams for each group of stars defined in Table \ref{tab:flares_types}. The Pleiades stars are marked by orange, the Hyades by black, and Praesepe by blue dots. The dashed lines mark the FFD fits above the completeness thresholds}.
    \label{fig:ffds}
\end{figure*}

\begin{figure}
    \centering
    \includegraphics[width=1\linewidth]{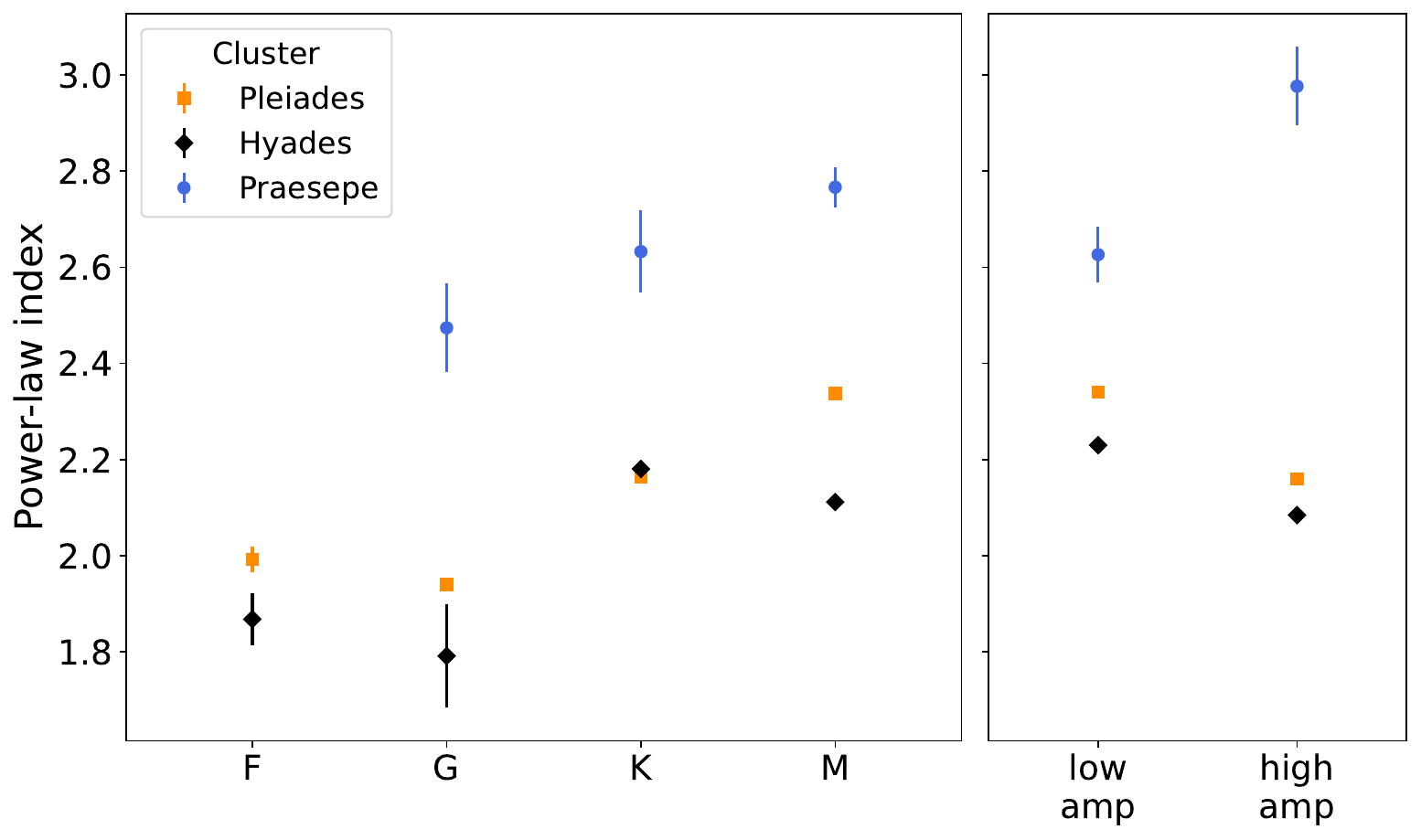}
    \includegraphics[width=1\linewidth]{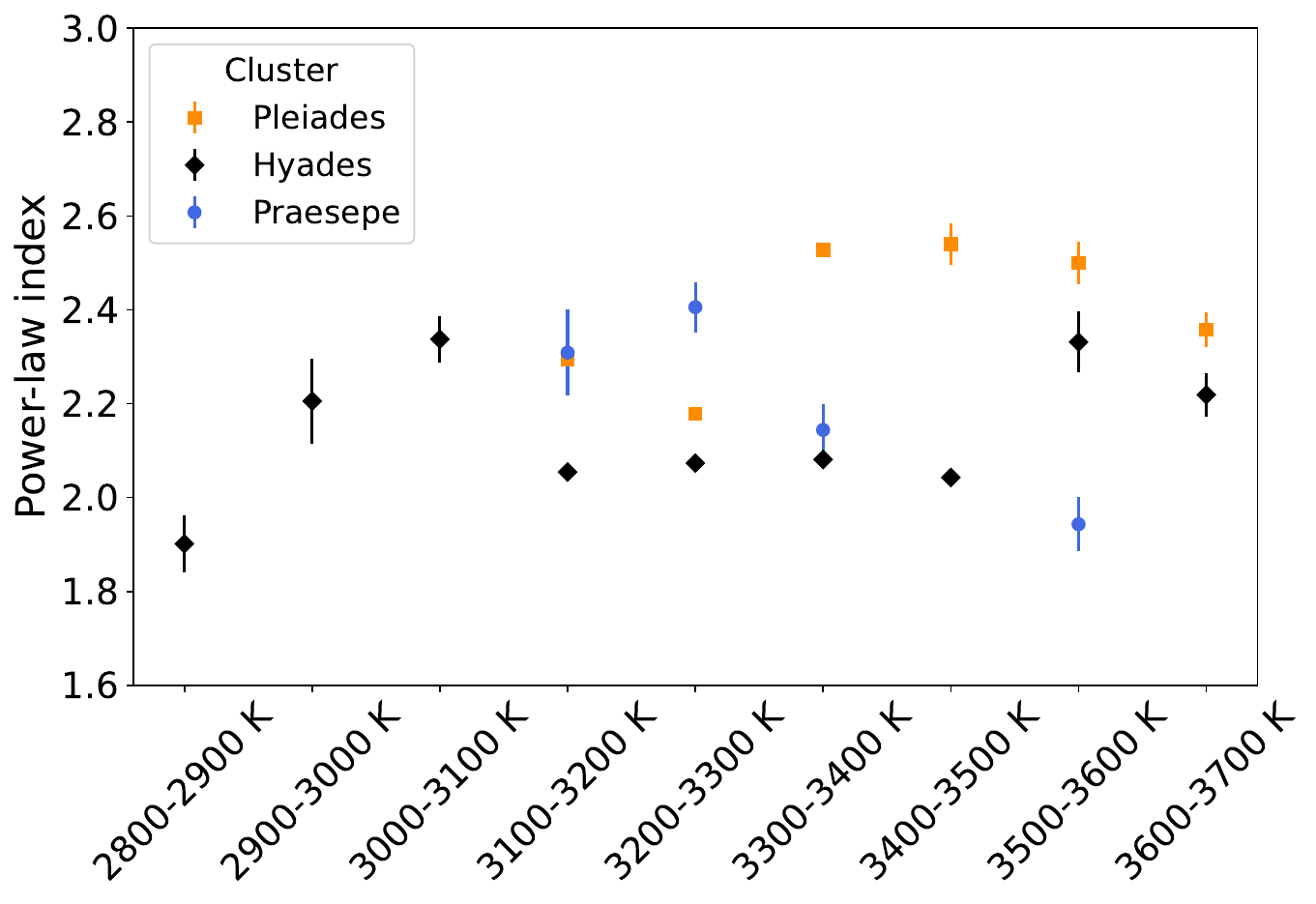}
    \caption{The top panel shows the power-law index ($\alpha_{\textrm{pl}}$) dependence on spectral type and light curve amplitude. The lower panel presents the power-law index dependence on effective temperature within the M spectral type.}
    \label{fig:slope}
\end{figure}

The FFDs corresponding to different spectral types and varying amplitudes are presented in Fig.~\ref{fig:ffds}. For F- and G-type stars the flare rate is higher in the Pleiades than in the Hyades by almost two orders of magnitude. For K-type stars, this difference decreases to roughly one order of magnitude, while the discrepancy becomes minimal for M-type stars. This variation can be assessed in the Praesepe cluster only for K- and M-type stars, as these are the spectral types for which relevant data are available. The occurrence rate for both spectral types remains below of what is observed in the Hyades cluster, although the difference is smaller for M-type stars than for K-types. Given that the Pleiades cluster has an age of approximately 100\,Myr, while the other two clusters are both around 700\,Myr old, this trend is consistent with the expectation that lower-temperature stars remain magnetically active over a longer period of time. The relatively low flare rate in Praesepe might be the consequence of the detection limit: since this is the farthest cluster of the three, we observe fewer of the faint, but typically more active  M dwarfs.

The power-law indices corresponding to the different groups are compared and presented in Fig. \ref{fig:slope}. In the Pleiades the FFD slope gradually becomes steeper from F-type stars toward M-type stars. A similar trend is observed in the Hyades between F- and K-type stars, although for the M-type group the slope becomes flatter again compared to that of the K-type stars. Praesepe exhibits the steepest FFDs across all groups, and the trend of the change in slope value matches that of the Pleiades cluster. When comparing the slopes of the high- and low-amplitude groups, it can be concluded that, stars exhibiting larger brightness variations tend to produce steeper FFDs.

Taking the analysis a step further, we divided the stars---and the corresponding flares---into temperature bins with 100\,K range, the coolest being the $2800-2900$\,K bin. We repeated the process for all clusters, and derive the FFDs for all bins with more than 3 stars. F-types have too few flares in all of the bins. With the exception of a few bins in the Pleiades, the same is largely true for the G stars. K- and M-type stars can be reasonably binned for the Pleiades and the Hyades, while for Praesepe it is only true for M dwarfs.

\begin{figure*}
    \centering
    \includegraphics[width=0.95\linewidth]{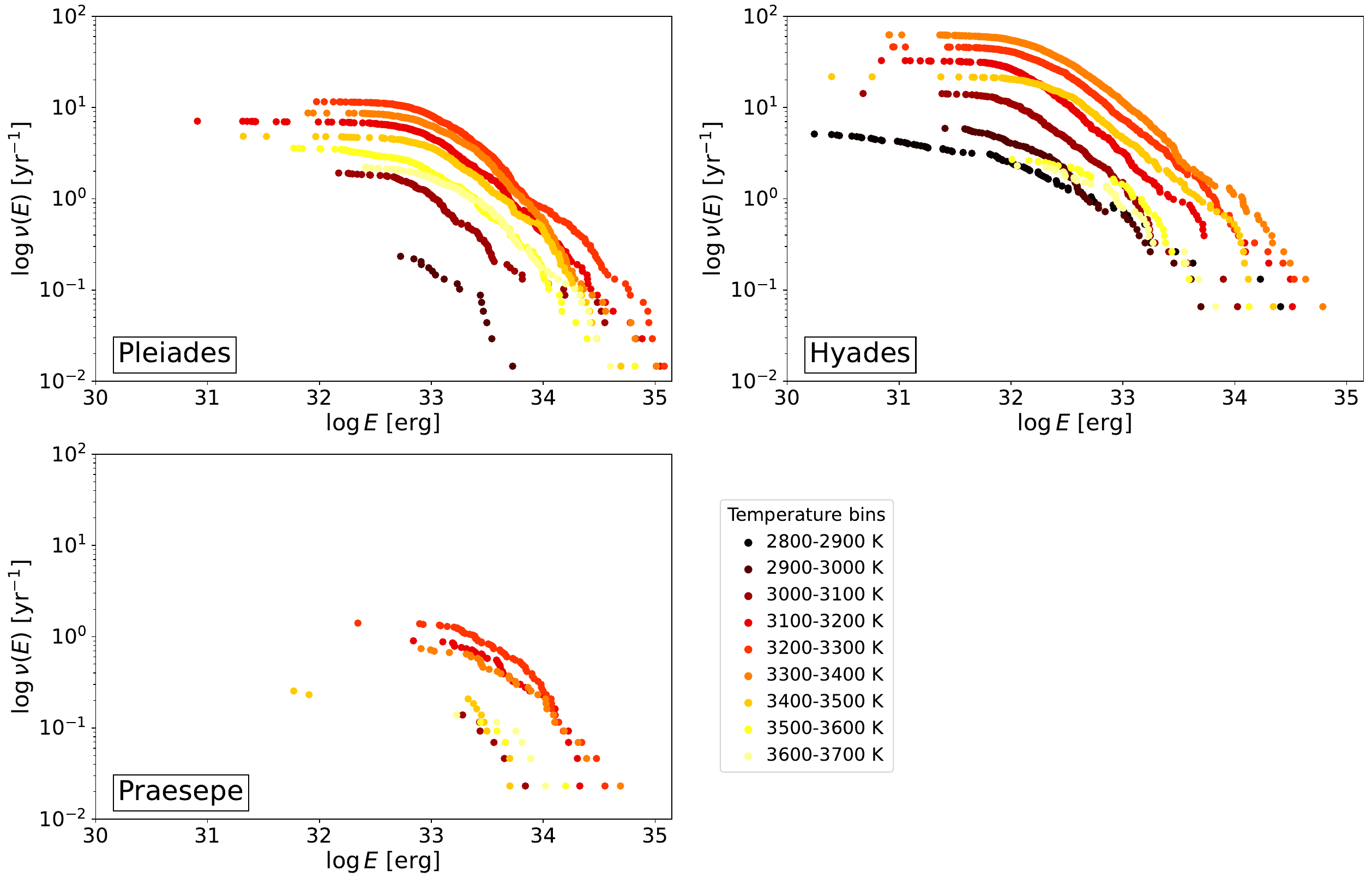}
    \caption{FFDs for 100 K temperature bins ranging from 2800 K to 3700 K. Starting from the coolest of stars, the flare rate grows smoothly until ~3200 K. At even higher temperatures the flare rate starts to gradually drop.}
    \label{fig:Mbins}
\end{figure*}

M stars show a smooth change in flare rates with the temperature (Fig.~\ref{fig:Mbins}): from the coolest ones to around 3200\,K the flare rate grows, from here toward higher temperatures it drops. This behavior is best visible for the Pleiades and the Hyades clusters. We found no similar change in the bins of the K-type stars (Fig.~\ref{fig:Kbins}). We also checked the variation of the power-law index of the temperature bin FFDs. The range of the fits varied only from cluster to cluster, not from one temperature bin to the other. The results are shown in Fig.~\ref{fig:slope}. The $\alpha_{\textrm{pl}}$ values range from $1.8$ to $2.6$, with no consistent trend observed across the clusters.

\section{Spectroscopic analysis}\label{sect:spectroscopy}

Figures \ref{fig:ple_spec} and \ref{fig:pra_spec} show dynamic spectra of the H$\alpha$ line and its environment for the stars of the Pleiades and Praesepe, respectively. We note that only those spectra are shown where the H$\alpha$ line had identifiable emission or absorption. Details of all stars with observed spectra are listed in Table~\ref{tab:spec}.

\begin{table}[]
\caption{The table contains the TIC identifier of the stars for which we obtained dynamic spectra, their effective temperature and whether the H$\alpha$ line is in emission (E), absorption (A) or not visible (---).}
\label{tab:spec}
\centering
\begin{tabular}{cccccc}
\hline \hline
\multicolumn{3}{c}{Pleiades}                            & \multicolumn{3}{c}{Praesepe}       \\ 
TIC       & $T_{\text{eff}}$ [K] & H$\alpha$  & TIC       & $T_{\text{eff}}$ [K] & H$\alpha$  \\ \hline
385509109 &  4232    & A & 175194928 & 3301    & A \\
385552305 & 3354    & A & 175233993 & 3800    & A \\
385589475 & 3445    & A & 175233949 & 3500    & A \\
385589362 & 3447    & E   & 175234031 & 4997    & E  \\
385509159 & 3363    & ---          & 307733505 & 3700    & E   \\
385552421 & 3319    & ---          &           &           &            \\
385589388 & 4565    & ---          &           &           &            \\
385552440 & 3423    & ---          &           &           &           \\
\hline 
\end{tabular}
\end{table}

Among the 11 observed Pleiades stars, three showed H$\alpha$ absorption and one emission, and four had no identifiable H$\alpha$ line.  From the original 8 Praesepe stars, three exhibit H$\alpha$ absorption and two emission. The recorded spectra for these stars were above S/N$=$10. Two stars from the Pleiades and three from the Praesepe were too faint for detection (with S/N below 4). One star from the Pleiades had to be discarded as it was near the edge of the FoV and part of its spectrum was not recorded.

None of these spectra display the characteristic signs of a CME, that would be a temporary enhancement of the blue wing of the line, corresponding to the ejected material. Co-temporal TESS observations are available for the stars of Praesepe in Sector 72 with 20\,s cadence. We manually checked these for flares, but regrettably no such event occurred over the duration of the spectroscopic observation time. The only candidate was observed on TIC 175194919 (around $\mbox{TBJD} = 3269.9409$), but closer inspection reveals cosmic ray hits marked with the \texttt{ApertureCosmic} quality flag.

\subsection{H$\alpha$ line variations}

Following \cite{h_alpha2003} we measured the H$\alpha$ chromospheric activity indices ($I_{\text{H}\alpha}$) for each spectrum in order to quantify the temporal variations in the line strength. The index is defined as the ratio of the mean flux within the line core to the average continuum flux from two reference bands: 

\begin{equation}
\hspace{0.35\linewidth}
I_{\text{H}\alpha}=\frac{\overline{F}_{\text{H}\alpha}}{0.5 \cdot (\overline{F}_1 +\overline{F}_2)}.
\end{equation}
$\overline{F}_{\text{H}\alpha}$ denotes the mean spectral flux in the $[-100, +100 ]\,\text{km}\,\text{s}^{-1}$ radial velocity interval centered on the core of the $6562.808$\,\AA\, $\text{H}\alpha$ line. $\overline{F}_1$ and  $\overline{F}_2$ are the mean fluxes from the $[-700,-300 ]\,\text{km}\,\text{s}^{-1}$ and $[+600,+1000]\,\text{km}\,\text{s}^{-1}$ intervals, respectively. As a consequence of its definition, the obtained indices are larger than one for emission lines and lower than that for absorption. 

\begin{figure*}[h!]
\centering
\begin{tabular}{cc}
\includegraphics[width=0.45\linewidth]{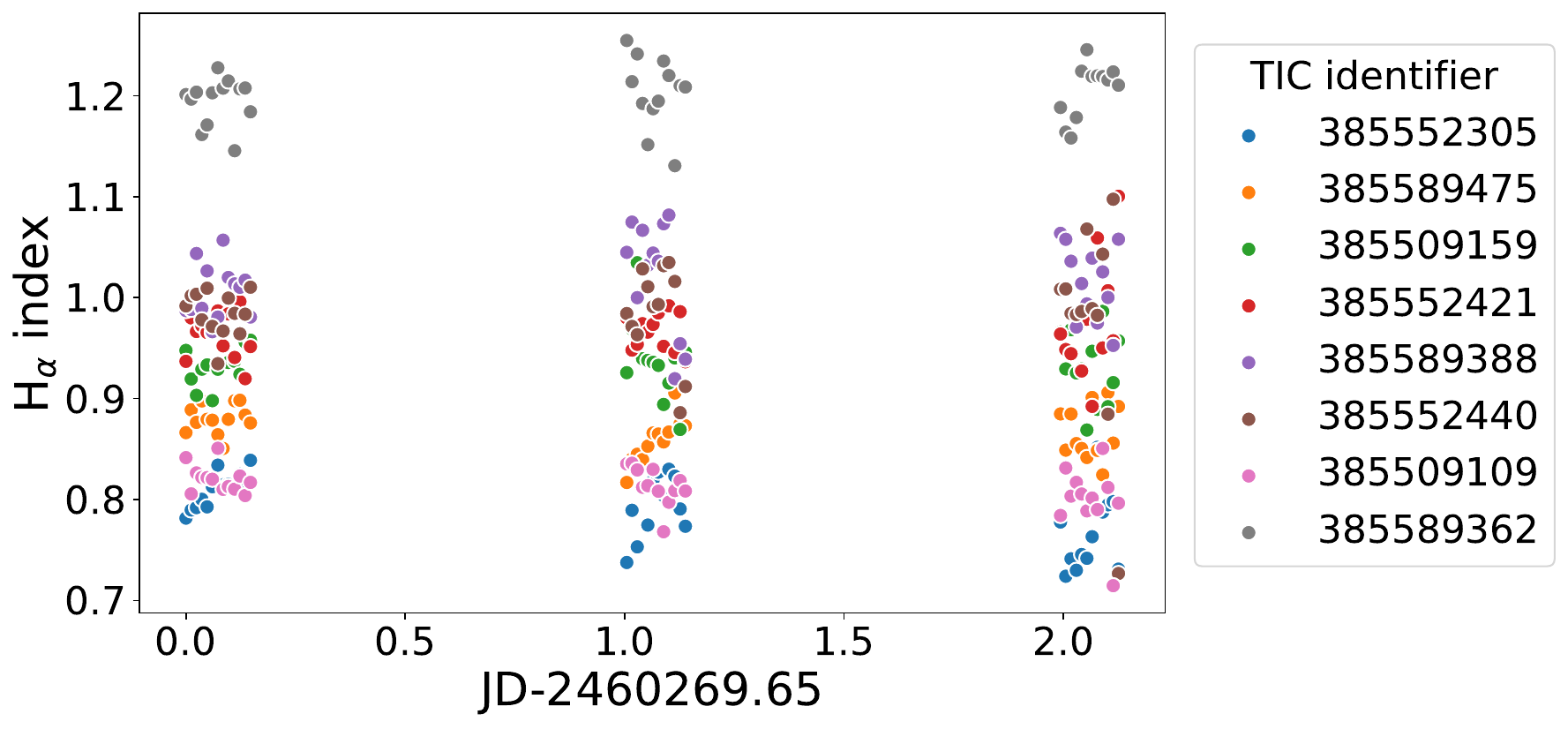}&
\includegraphics[width=0.45\linewidth]{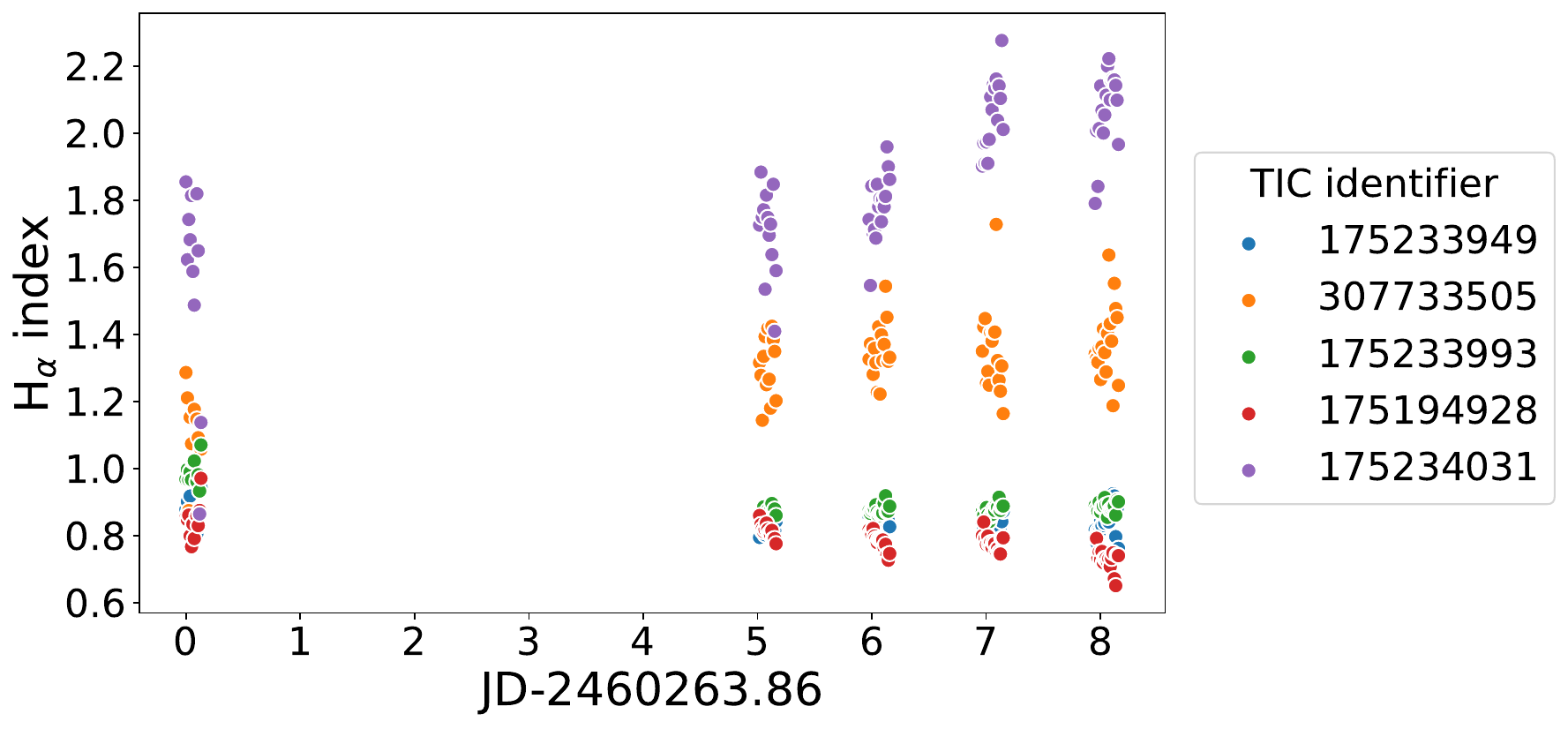}
\end{tabular}
\caption{The behavior of H$\alpha$ indices in time, their values are larger than one for emission lines and lower than that for absorption. The left panel shows the Pleiades stars, while the right panel shows the Praesepe stars.}
\label{fig:ha_vs_time}

\end{figure*}

\begin{figure}[h!]
    \centering
    \includegraphics[width=1\linewidth]{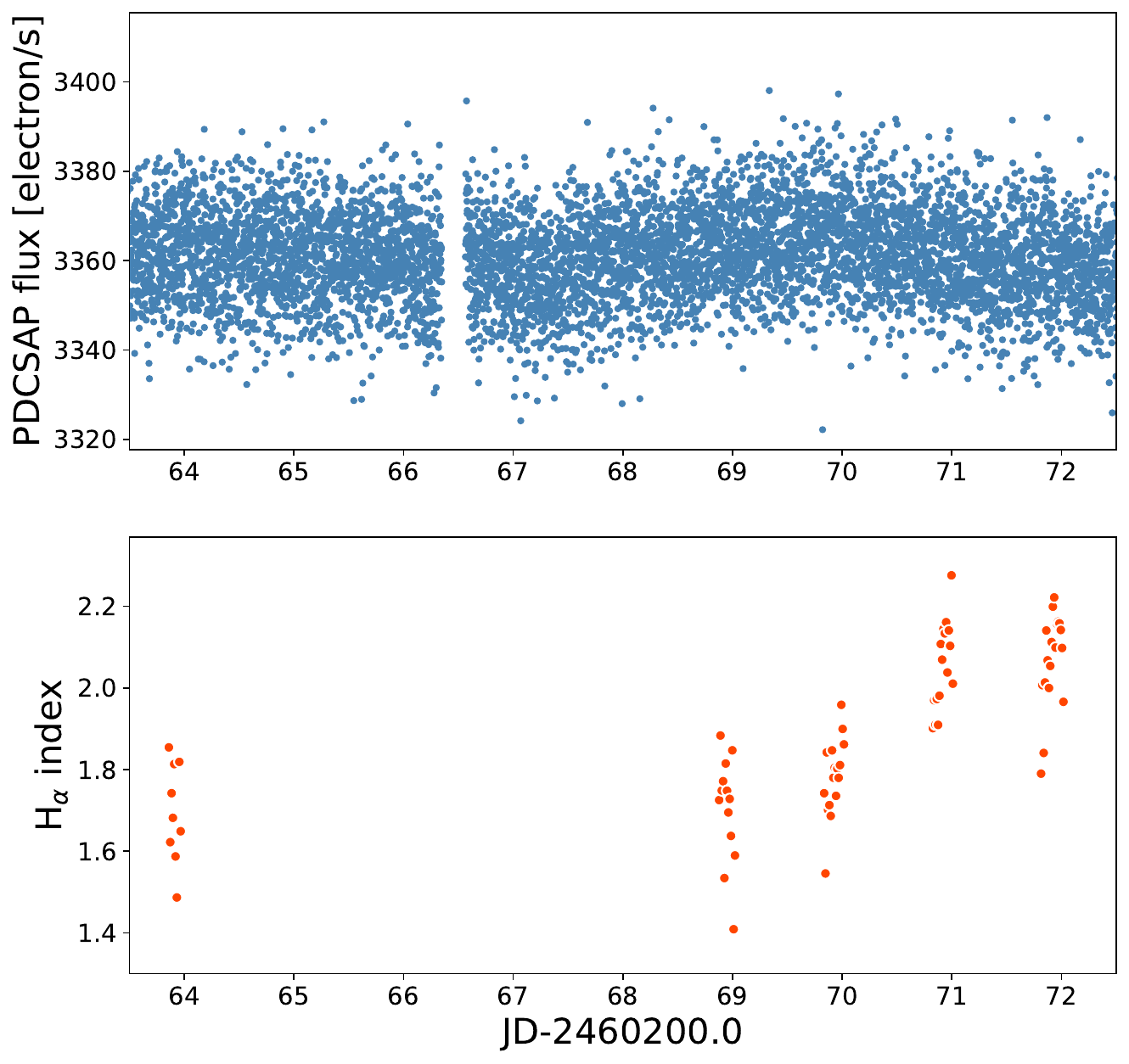}
    \caption{H$\alpha$ indices of TIC 175234031 from the Praesepe cluster and its co-temporal TESS light curve.}
    \label{fig:tic031}
\end{figure}

The temporal variations of the index values for each star with H$\alpha$ emission or absorption are shown in Fig.~\ref{fig:ha_vs_time}. The Pleiades stars show no significant variation and the same is true for most of the Praesepe stars. The exception is TIC 175234031 where the emission line became stronger for the last two observing nights. Since this is one of the stars with co-temporal TESS observations, we compare the line variation to that of the light curve (Fig.~\ref{fig:tic031}). At the time of the enhanced line emission the star is in a fainter state in the TESS band, indicating that dark active regions are in view. This agrees with previous results \citep[e.g.,][]{Namekataetal2022a} that the heightened chromospheric activity traced by the strengthening of the line corresponds to stellar spots.

\section{Discussion}\label{sect:discussion}

\subsection{Flare activity}

Previous studies suggest that a relationship between flaring, temperature and age exists for active stars.
The studies by \citet{2019A&A...622A.133I,2021A&A...645A..42I} using Kepler/K2 data include all three open clusters examined in our work, providing a useful benchmark for comparison. Direct comparison between Kepler/K2- and TESS-derived flare properties should, however, be made with some caution. The different instrumental bandpasses, cadences, photometric precision, and observing strategies affect flare detection completeness, especially for the lower energy range. In particular, the redder TESS bandpass generally yields lower observed flare amplitudes than the Kepler bandpass for the same flare \citep{2019MNRAS.489..437D}. This increases the minimum amplitude that is required for the flares to be detectable. Nevertheless, the high-energy part of the flare frequency distribution, from which the power-law slope $\alpha_{\rm pl}$ is usually derived, is expected to be much less sensitive to these instrumental differences, allowing meaningful comparison of the FFDs derived from the different missions. \citet{2019A&A...622A.133I} investigated the activity of three open clusters (M44, M45, M67) with solar metallicity using data detrended with the K2 Systematics Correction (K2SC) pipeline \citep{2016MNRAS.459.2408A} from the Kepler/K2 missions. Their sample mostly included late-K and early to mid-M dwarfs, adding up to 1761 target stars. Their analysis found that the slopes of the FFDs for stars grouped by temperature ranged from 2.0 to 2.4 (generally consistent with our result) and it showed no obvious trend with mass or age. Although the FFD-slope did not seem to depend on age, other activity indicators—such as average flare frequency—did show a dependence.
\cite{2021A&A...645A..42I} found $\alpha_{\textrm{pl}} \approx 1.84-2.39 $ based on flares from long cadence light curves of the Kepler/K2 mission detrended with the K2SC pipeline. They investigated the stellar activity in five open clusters (including all three of ours) with ages from zero age main sequence (ZAMS) to 3.6\,Gyr. Using effective temperature bins, they found that flaring activity decreases from mid-M dwarfs to G stars and from ZAMS to solar age. The exception to this were stars later than M5, here they found the activity to be increasing from ZAMS to 690\,Myr. They also found evidence for a mass- and rotation-dependent threshold age beyond which stars leave the saturated flaring activity regime. 

\cite{2024AJ....168...60F} measured the flare rates for stars younger than 300\,Myr. Their sample stars were members of 26 nearby, young moving groups, associations, and open clusters, and their analysis was executed using TESS 2 minutes cadence simple aperture photometry (SAP) light curves. They found that $\alpha_{\textrm{pl}}$ saturates at $\alpha_{\textrm{pl}}$ = 1.2 to 1.6 for ages younger than 300\,Myr and shows a decline for older stars.

\cite{2026ApJ...999..252T} investigated stars that belong to nearby young moving groups aged 10--150\,Myr based on TESS Quick Look Pipeline \citep{2020RNAAS...4..204H} light curves. They found flaring activity to be universal in this age range with no clear dependence on age for any spectral type. They identified M-types to have flare frequencies 5 times higher than those of G--K type stars, and an order of magnitude higher than F-type stars. Their FFDs yielded a power law index of $\alpha_{\textrm{pl}} = 1.51$.

\cite{2019ApJS..241...29Y} analyzed long cadence SAP light curves of the Kepler mission to construct a large catalog of stellar flares containing more than $162\,000$ flare events from 3420 flare stars . Their study showed that flare activity becomes much stronger toward cooler stars, particularly K and M dwarfs, and that rapidly rotating stars tend to be more magnetically active. They found that flare frequency distributions follow a power-law relation, with indices of  $1.8-2.0$, indicating a broadly common magnetic reconnection mechanism from F-type stars to M-types.

Our analysis yielded power-law indices in the $1.8-2.8$ range which is broadly consistent with previous studies described above. There appears to be no obvious connection between $\alpha_{\textrm{pl}}$ and the age of the discussed clusters, however the FFDs of K- and M-type stars are steeper in all cases than that of F- and G-stars. In agreement with \cite{2019ApJS..241...29Y} we found that G-type stars produce the least steep FFDs. 

In contrast to the $\alpha_{\textrm{pl}}$ values, the flare rates exhibit a clear dependence on cluster age for different spectral types. Among the three investigated clusters, the Pleiades consistently displays the highest flare rates across all spectral classes, in agreement with its status as the youngest cluster in the sample (see Fig. \ref{fig:ffds}).
M-type stars in the Hyades show flare rates comparable to those observed for M dwarfs in the Pleiades, indicating that magnetic activity in low-mass stars remains efficient even at the older cluster age. In contrast, K-type stars exhibit a pronounced decline in flare rate between the two clusters. This behavior is consistent with the more rapid rotational spin-down of K dwarfs relative to M dwarfs \citep[e.g.,][]{2003ApJ...586..464B,2011ASPC..448..841D}, which leads to a faster reduction in magnetic activity and, consequently, flare production \citep[e.g.,][]{2007ApJ...666..393C,2019ApJ...871..241D,2022ApJ...931...45N}. These results suggest that the $\sim$500\,Myr age difference between the two clusters is sufficient for K-type stars to undergo a substantial decrease in flare rate, whereas M-type stars preserve elevated levels of magnetic activity over the same evolutionary timescale.

\subsection{Dynamo mechanisms of M-type stars}

\cite{2018ApJ...861L..11J} reported the discovery of a narrow, well-defined under-density in the lower main sequence of the Hertzsprung–Russell diagram using high-precision photometry and parallaxes from Gaia Data Release 2. The gap appears among nearby M-dwarf stars around the effective temperatures of roughly 3200–3300 K (spectral type M3V) and is observed consistently across different datasets and wavelength bands, indicating it is not an observational artifact. 
The authors interpreted this gap as evidence of the transition to a fully convective state in low-mass stars, reflecting a change in internal stellar structure that affects their luminosity--color relation.

We found that the flare rate of M-type stars varies smoothly with effective temperature: it increases from $\sim$2800\,K to $\sim$$3200-3300$\,K (see Fig.~\ref{fig:Mbins}), and decreases at higher temperatures. This temperature range coincides with the transition from fully convective to partially convective stars reported by \cite{2018ApJ...861L..11J}.

\cite{2018MNRAS.480.1711M} explain the narrow gap found by \cite{2018ApJ...861L..11J} as a consequence of standard stellar evolution rather than transition from fully to partially convective stars. They explain that in stars around $\sim 0.3 -0.35$ solar masses (M$_\odot$), the interior develops and later erases a thin radiative zone as $^3$He accumulates and is mixed when the core and envelope convection zones merge, briefly altering the structure of the stars and energy generation. This causes stars in that mass range to evolve more rapidly through a small range of luminosities, producing a slight dip in the stellar luminosity function. When a realistic population of stars is simulated, this dip appears as the observed under-density in the Gaia HRD. Based on their results we can determine that by the age of the Pleiades ($\sim$\,100\,Myr) the erasure of the radiative zone happens for $0.3$ M\,$_\odot$ stars. At the age of the Hyades and Praesepe ($\sim$\,700\,Myr) the previous is true for stars with $0.3-0.31$ M\,$_\odot$. These masses correspond to $T_{\rm eff}\sim$3200\,K, the same temperature where we found the peak of the flare rate.

\subsection{Color--rotation diagram}

The activity of stars---and therefore their flaring---is closely connected to their rotation, which in turn depends on their age. Color--rotation diagrams (CRD) are widely used in gyrochronology \citep{2003ApJ...586..464B} to study the age--rotation relationship based on open cluster stars. We adopted the rotational periods from \cite{2020ApJ...904..140C} and  binarity from \cite{2025AJ....169..116L} for the Pleiades and both properties from \cite{2022ApJ...931...45N} for the Hyades and Praesepe and plotted their CRDs colored by the flare rate of the stars in Fig.\,\ref{fig:prot_vs_color3}. By the time the stars reach the age of the Hyades and Praesepe, the stars on the I sequence (diagonal pattern at $\sim$10\,days period, marks stars with interface dynamo) do not produce flares with high enough energies to be detectable by TESS. However, this should not be interpreted as implying that stars on the I sequence are flare-inactive, but rather that lower-energy flares may remain below the TESS detection threshold. Indeed, \citet{2019ApJS..241...29Y} detected flaring events in stars already evolved onto the unsaturated (I-sequence) regime. On the diagonal part of the I sequence the rotation of the convective zone has reached that of the core of the stars. As expected, the I sequence on the Pleiades CRD deviates from the diagonal, as the cluster is only $\sim$100\,Myr old and its members are still rotating faster. We can infer that the energy of the produced flares drop when the core and the convective zone reach the same rotational speed and therefore magnetically couple.

Recent studies have suggested that active main-sequence stars in intermediate-age clusters such as Praesepe and NGC~6811 \citep{2023ApJ...951L..49C,2025A&A...697A.177S} may experience a phase of spin-down stalling, during which magnetic braking becomes temporarily less efficient. Within the framework of the rotation–activity relation, such stars are expected to maintain elevated magnetic activity and flare rates for longer than predicted by standard spin-down models. Although the higher flare rates observed among the rapidly rotating Praesepe members (bottom panel of Fig.\,\ref{fig:prot_vs_color3}) are qualitatively consistent with this picture, our data do not allow us to isolate the effect of spin-down stalling from the general dependence of magnetic activity on rotation.

\subsection{Dynamic spectra}

Out of the 13 observed dynamic spectra seven exhibited H$\alpha$ absorption, three H$\alpha$ emission lines while the rest shows neither. Two of the stars with emission line are M dwarfs (TIC~307733505 and TIC~385589362) while one (TIC~175234031) is a K-type star. In inactive late-type stars H$\alpha$ is typically in absorption as a consequence of their cool, magnetically weakly active chromospheres.
Earlier studies by \cite{1979ApJ...234..579C} and \cite{1986ApJS...61..531S} interpreted weak or absent H$\alpha$ signatures observed in some M stars as evolutionary changes in stellar magnetic and chromospheric activity. In this scenario, the gradual filling-in of the photospheric H$\alpha$ absorption profile by the emission line from the chromosphere---originating from numerous active regions---signal the transition between an inactive M star and a relatively active one.

\subsection{CME rate}

Based on the non-detections of CMEs during the spectral observations we can give an upper limit on the frequency of events. Using Fig.~1 of \cite{2020MNRAS.494.3766O} we can expect the detection of CMEs with masses above $\sim 10^{17}$\,g using our MDM/OSMOS spectra of K--M stars with an average $S/N \approx 15$. If we assume $k=0$ CME events were observed during an observation time $T$, the Poisson event rate $\lambda$ (events per day) can be constrained as follows.

We use Bayes theorem to calculate the posterior probability density function $\mathcal{P}(\lambda \mid k=0)$ by multiplying the Poisson likelihood $\mathcal{L}(k=0 \mid \lambda) = e^{-\lambda T}$ with the prior distribution $\pi(\lambda)$. As the likelihood is not strongly constraining in this case, we need to choose the prior carefully. We consider four different options: a flat uniform prior $\pi(\lambda) \propto 1$ encoding no prior information (similar to \citealt{2020MNRAS.494.3766O}); the scale invariant Jeffreys prior $\pi(\lambda) \propto \lambda^{-1/2}$, which is invariant under reparameterization; and weakly informative priors encoding the $\lambda_0$ flare rate observed by TESS using a lognormal distribution $\pi(\lambda) \propto \mathrm{LogNorm}(\mu=\lambda_0,\sigma=2)$, where $\mu$ is the location and $\sigma$ is the scale parameter; and a Gamma distribution $\pi(\lambda) \propto \mathrm{Gamma}(\alpha=1,\beta=\lambda_0)$, where $\alpha$ is the shape and $\beta$ is the rate parameter. In the latter two cases, we assume that every flare above a chosen energy threshold is connected to a detectable CME, so we can place a prior on the spectroscopic CME rate based on photometric data. We selected the parameters of the weakly informative priors to center around $\lambda_0$, and to allow for a large range of values: $\mathrm{LogNorm}(\mu=\lambda_0,\sigma=2)$ allows for roughly two orders of magnitude variations, and $\mathrm{Gamma}(\alpha=1,\beta=\lambda_0)$ flattens for small values. Thus, the posterior distribution using the different priors follows as:

\begin{equation}
\mathcal{P}(\lambda \mid 0) \propto
\begin{cases}
e^{-\lambda T}
& \text{flat} \\

\lambda^{-1/2} e^{-\lambda T}
& \text{Jeffreys} \\

\mathrm{LogNorm}(\mu, \sigma) e^{-\lambda T}
& \text{weakly informative} \\

\mathrm{Gamma}(\alpha, \beta) e^{-\lambda T}
& \text{weakly informative}
\end{cases}
\end{equation}

We then use these posteriors to calculate the 95\% credible upper limit for the CME rate ($\lambda_{95}$) numerically as:

\begin{equation}
\label{eq:upper_limit_integral}
    \int_0^{\lambda_{95}} \mathcal{P}(\lambda \mid 0)\,d\lambda = 0.95.
\end{equation}

\begin{figure}[h!]
    \centering
    \includegraphics[width=\linewidth]{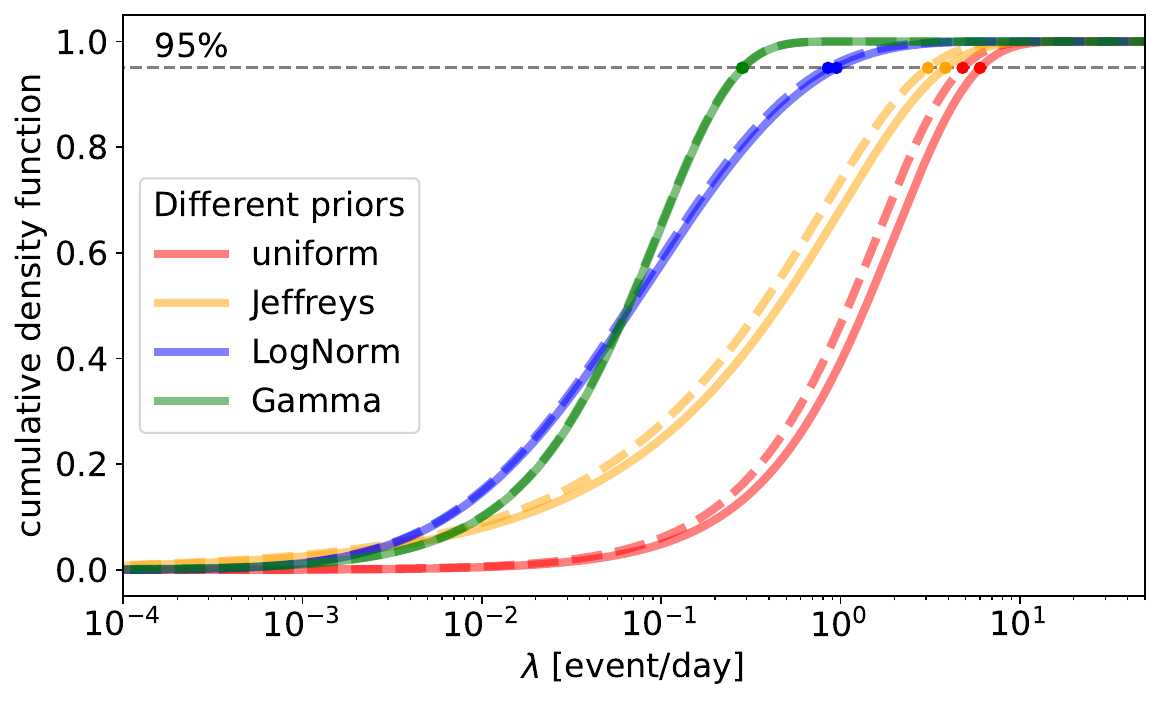}
    \caption{Cumulative posterior distributions for the CME rate for the stars observed by MDM/OSMOS in the Pleiades with solid lines, and Praesepe with dashed lines. Colors indicate the posteriors calculated with different priors, and the colored points denote the 95\% credible upper limits.}
    \label{fig:cme_rate_posterior}
\end{figure}

Based on solar flares analyzed by \cite{2013ApJ...764..170D}, we can expect a CME with mass of $10^{17}$\,g to be accompanied by a flare with $\sim$$10^{31}$\,erg energy in the Geostationary Operational Environmental Satellite \citep[GOES,][]{1994SoPh..154..275G} X-ray band, which corresponds to roughly $\sim$$10^{32}$\,erg in the TESS band, based on the ratio of bolometric to GOES band energies of solar flares from \cite{2012ApJ...759...71E}, and converting that to the TESS band by assuming a blackbody emission with 10000\,K temperature. Using the TESS FFDs of the stars with spectroscopic observations from Table~\ref{tab:spec}, we expect on average $\lambda_0 \approx 0.1$\,flares/day with energies above $10^{32}$\,erg. We use this $\lambda_0$ value in the weakly informative priors for all stars, and $T=12$\,h of observing time for the Pleiades and $T=15$\,h for Praesepe. The results are summarized in Table~\ref{tab:cme_rates}, and the cumulative posterior distributions are shown in Fig.~\ref{fig:cme_rate_posterior}. We note that these results are moderately sensitive to the prior hyperparameters ($\sim20$\% change in $\lambda_{95}$ for $\sigma=1.5-4$ for the lognormal prior, and $\alpha=0.5 - 2$ for the Gamma prior). Based on this analysis, we can exclude CME rates above a few events per day for these stars. The limits become more constraining when we incorporate prior information about the flare rate.

\begin{table}[h]
\caption{95\% credible upper limits for the CME rate of individual stars from the Bayesian analysis.}
\label{tab:cme_rates}
\centering
\begin{tabular}{ccc}
\hline \hline
 & $\lambda_{\text{Pleiades}}$ & $\lambda_{\text{Praesepe}}$ \\
 & [d$^{-1}$] & [d$^{-1}$] \\
 \hline
uniform prior & $<6.0$ & $<4.8$ \\
Jeffreys prior & $<3.9$ & $<3.1$ \\
LogNorm prior & $<1.0$ & $<0.9$ \\
Gamma prior & $<0.3$ & $<0.3$ \\
\hline
\end{tabular}
\end{table}

\section{Conclusions}\label{sect:conclusions}

We used TESS photometry and MDM/OSMOS spectroscopic data from three open clusters: the Pleiades, the Hyades and Praesepe. In the following, we summarize our main conclusions from investigating the magnetic activity of their member stars.

\begin{itemize}

\item The power-law indices obtained in our analysis span the $1.8-2.8$ range and are broadly consistent with earlier studies --- these show no clear dependence on cluster age.

\item We found higher flare rates in the Pleiades than in the Hyades and Praesepe for F-,G-, and K-type stars. It matches the expectation that younger stars have higher flare rates.

\item The flare rate of M dwarfs is similar in the Pleiades and the Hyades while for K-type stars the flare rate in the Pleiades is an order of magnitude higher. The reason for the difference is that K stars in the Hyades have already spun down as opposed to the Pleiades.

\item Our study shows that the flare rate in M dwarfs changes with their effective temperature. It is highest for $T_{\rm eff}\sim 3200-3000$\,K while moving to cooler and hotter M dwarfs the flare rate smoothly decreases. This temperature marks either the upper limit of the fully convective stars \citep{2020ApJ...891..128M} or the transition from it \citep{2018ApJ...861L..11J}.

\item We collected MDM/OSMOS low-resolution spectra of cluster members to search for CME signatures in the H$\alpha$ line, but we detected no such signal. Using Bayesian analysis, we converted this non-detection to a 95\% credible upper limit on the CME rate, and find $<0.3-6$\,d$^{-1}$, depending on the prior.

\end{itemize}

Our results are generally consistent with earlier studies of flare statistics in young open clusters and nearby field stars. Our analysis also provides additional support for the classical age--activity relation.

\begin{acknowledgements}

The authors thank the anonymous referee for improving the quality of the paper with helpful comments and suggestions. Funding for this research was provided by the Hungarian National Research, Development and Innovation Office grant \'Elvonal KKP-143986. 
As part of the \textit{"Looking for stellar CMEs on different wavelengths"} project we are grateful for the possibility of using HUN-REN Cloud \cite{MTACloud} which helped us achieve the results published in this paper.
This paper includes data collected by the TESS mission. Funding for the TESS mission is provided by the NASA's Science Mission Directorate.
This work is based on observations obtained at the MDM Observatory, operated by Dartmouth College, Columbia University, Ohio State University, Ohio University, and the University of Michigan. 

\end{acknowledgements}

\bibliographystyle{aa}
\bibliography{bib}

@ARTICLE{tess,
       author = {{Ricker}, George R. and {Winn}, Joshua N. and {Vanderspek}, Roland and {Latham}, David W. and {Bakos}, G{\'a}sp{\'a}r {\'A}. and {Bean}, Jacob L. and {Berta-Thompson}, Zachory K. and {Brown}, Timothy M. and {Buchhave}, Lars and {Butler}, Nathaniel R. and {Butler}, R. Paul and {Chaplin}, William J. and {Charbonneau}, David and {Christensen-Dalsgaard}, J{\o}rgen and {Clampin}, Mark and {Deming}, Drake and {Doty}, John and {De Lee}, Nathan and {Dressing}, Courtney and {Dunham}, Edward W. and {Endl}, Michael and {Fressin}, Francois and {Ge}, Jian and {Henning}, Thomas and {Holman}, Matthew J. and {Howard}, Andrew W. and {Ida}, Shigeru and {Jenkins}, Jon M. and {Jernigan}, Garrett and {Johnson}, John Asher and {Kaltenegger}, Lisa and {Kawai}, Nobuyuki and {Kjeldsen}, Hans and {Laughlin}, Gregory and {Levine}, Alan M. and {Lin}, Douglas and {Lissauer}, Jack J. and {MacQueen}, Phillip and {Marcy}, Geoffrey and {McCullough}, Peter R. and {Morton}, Timothy D. and {Narita}, Norio and {Paegert}, Martin and {Palle}, Enric and {Pepe}, Francesco and {Pepper}, Joshua and {Quirrenbach}, Andreas and {Rinehart}, Stephen A. and {Sasselov}, Dimitar and {Sato}, Bun'ei and {Seager}, Sara and {Sozzetti}, Alessandro and {Stassun}, Keivan G. and {Sullivan}, Peter and {Szentgyorgyi}, Andrew and {Torres}, Guillermo and {Udry}, Stephane and {Villasenor}, Joel},
        title = "{Transiting Exoplanet Survey Satellite (TESS)}",
      journal = {Journal of Astronomical Telescopes, Instruments, and Systems},
         year = 2015,
        month = jan,
       volume = {1},
          eid = {014003},
        pages = {014003},
          doi = {10.1117/1.JATIS.1.1.014003},
       adsurl = {https://ui.adsabs.harvard.edu/abs/2015JATIS...1a4003R}
}

@inproceedings{10.1117/12.2233418,
    author = {Jon M. Jenkins and Joseph D. Twicken and Sean McCauliff and Jennifer Campbell and Dwight Sanderfer and David Lung and Masoud Mansouri-Samani and Forrest Girouard and Peter Tenenbaum and Todd Klaus and Jeffrey C. Smith and Douglas A. Caldwell and A. Dean Chacon and Christopher Henze and Cory Heiges and David W. Latham and Edward Morgan and Daryl Swade and Stephen Rinehart and Roland Vanderspek},
    title = {{The TESS science processing operations center}},
    volume = {9913},
    booktitle = {Software and Cyberinfrastructure for Astronomy IV},
    editor = {Gianluca Chiozzi and Juan C. Guzman},
    organization = {International Society for Optics and Photonics},
    publisher = {SPIE},
    pages = {99133E},
    year = {2016},
    doi = {10.1117/12.2233418},
    URL = {https://doi.org/10.1117/12.2233418}
}

@article{MTACloud,
	doi = {10.22503/inftars.xxii.2022.2.8},
	url = {https://doi.org/10.22503%2Finftars.xxii.2022.2.8},
	year = 2022,
	month = {aug},
	publisher = {Informacios Tarsadalom},
	volume = {22},
	number = {2},
	pages = {128},
	author = {Mih{\'{a}}ly H{\'{e}}der and Ern{\H{o}} Rig{\'{o}} and Dorottya Medgyesi and R{\'{o}}bert Lovas and Szabolcs Tenczer and Ferenc Török and Attila Farkas and M{\'{a}}rk Em{\H{o}}di and J{\'{o}}zsef Kadlecsik and György Mez{\H{o}} and {\'{A}}d{\'{a}}m Pint{\'{e}}r and P{\'{e}}ter Kacsuk},
	title = {The Past, Present and Future of the {ELKH} Cloud},
	journal = {Inform{\'{a}}ci{\'{o}}s T{\'{a}}rsadalom}
}

@ARTICLE{2021A&A...645A..42I,
       author = {{Ilin}, Ekaterina and {Schmidt}, Sarah J. and {Poppenh{\"a}ger}, Katja and {Davenport}, James R.~A. and {Kristiansen}, Martti H. and {Omohundro}, Mark},
        title = "{Flares in open clusters with K2. II. Pleiades, Hyades, Praesepe, Ruprecht 147, and M 67}",
      journal = {\aap},
         year = 2021,
        month = jan,
       volume = {645},
          eid = {A42},
        pages = {A42},
          doi = {10.1051/0004-6361/202039198},
archivePrefix = {arXiv},
       eprint = {2010.05576},
 primaryClass = {astro-ph.SR},
       adsurl = {https://ui.adsabs.harvard.edu/abs/2021A&A...645A..42I}
}

@ARTICLE{2026A&A...706A..54G,
       author = {{G{\"o}rgei}, A. and {Kriskovics}, L. and {Vida}, K. and {Seli}, B. and {Ol{\'a}h}, K. and {S{\'a}gi}, P. and {B{\'o}di}, A. and {J{\"a}rvinen}, S.~P. and {Strassmeier}, K.~G. and {P{\'a}l}, A. and {K{\H{o}}v{\'a}ri}, Zs.},
        title = "{Magnetic activity on the young Sun: A case study of EK Draconis}",
      journal = {\aap},
         year = 2026,
        month = jan,
       volume = {706},
          eid = {A54},
        pages = {A54},
          doi = {10.1051/0004-6361/202556949},
archivePrefix = {arXiv},
       eprint = {2512.03830},
 primaryClass = {astro-ph.SR},
       adsurl = {https://ui.adsabs.harvard.edu/abs/2026A&A...706A..54G}
}

@INPROCEEDINGS{nextgen,
       author = {{Allard}, F. and {Homeier}, D. and {Freytag}, B.},
        title = "{Model Atmospheres From Very Low Mass Stars to Brown Dwarfs}",
    booktitle = {16th Cambridge Workshop on Cool Stars, Stellar Systems, and the Sun},
         year = 2011,
       editor = {{Johns-Krull}, Christopher and {Browning}, Matthew K. and {West}, Andrew A.},
       series = {Astronomical Society of the Pacific Conference Series},
       volume = {448},
        month = dec,
        pages = {91},
          doi = {10.48550/arXiv.1011.5405},
archivePrefix = {arXiv},
       eprint = {1011.5405},
 primaryClass = {astro-ph.SR},
       adsurl = {https://ui.adsabs.harvard.edu/abs/2011ASPC..448...91A}
}

@ARTICLE{gaia_bolcorr,
       author = {{Creevey}, O.~L. and {Sordo}, R. and {Pailler}, F. and {Fr{\'e}mat}, Y. and {Heiter}, U. and {Th{\'e}venin}, F. and {Andrae}, R. and {Fouesneau}, M. and {Lobel}, A. and {Bailer-Jones}, C.~A.~L. and {Garabato}, D. and {Bellas-Velidis}, I. and {Brugaletta}, E. and {Lorca}, A. and {Ordenovic}, C. and {Palicio}, P.~A. and {Sarro}, L.~M. and {Delchambre}, L. and {Drimmel}, R. and {Rybizki}, J. and {Torralba Elipe}, G. and {Korn}, A.~J. and {Recio-Blanco}, A. and {Schultheis}, M.~S. and {De Angeli}, F. and {Montegriffo}, P. and {Abreu Aramburu}, A. and {Accart}, S. and {{\'A}lvarez}, M.~A. and {Bakker}, J. and {Brouillet}, N. and {Burlacu}, A. and {Carballo}, R. and {Casamiquela}, L. and {Chiavassa}, A. and {Contursi}, G. and {Cooper}, W.~J. and {Dafonte}, C. and {Dapergolas}, A. and {de Laverny}, P. and {Dharmawardena}, T.~E. and {Edvardsson}, B. and {Le Fustec}, Y. and {Garc{\'\i}a-Lario}, P. and {Garc{\'\i}a-Torres}, M. and {Gomez}, A. and {Gonz{\'a}lez-Santamar{\'\i}a}, I. and {Hatzidimitriou}, D. and {Jean-Antoine Piccolo}, A. and {Kontiza}, M. and {Kordopatis}, G. and {Lanzafame}, A.~C. and {Lebreton}, Y. and {Licata}, E.~L. and {Lindstr{\o}m}, H.~E.~P. and {Livanou}, E. and {Magdaleno Romeo}, A. and {Manteiga}, M. and {Marocco}, F. and {Marshall}, D.~J. and {Mary}, N. and {Nicolas}, C. and {Pallas-Quintela}, L. and {Panem}, C. and {Pichon}, B. and {Poggio}, E. and {Riclet}, F. and {Robin}, C. and {Santove{\~n}a}, R. and {Silvelo}, A. and {Slezak}, I. and {Smart}, R.~L. and {Soubiran}, C. and {S{\"u}veges}, M. and {Ulla}, A. and {Utrilla}, E. and {Vallenari}, A. and {Zhao}, H. and {Zorec}, J. and {Barrado}, D. and {Bijaoui}, A. and {Bouret}, J. -C. and {Blomme}, R. and {Brott}, I. and {Cassisi}, S. and {Kochukhov}, O. and {Martayan}, C. and {Shulyak}, D. and {Silvester}, J.},
        title = "{Gaia Data Release 3. Astrophysical parameters inference system (Apsis). I. Methods and content overview}",
      journal = {\aap},
         year = 2023,
        month = jun,
       volume = {674},
          eid = {A26},
        pages = {A26},
          doi = {10.1051/0004-6361/202243688},
archivePrefix = {arXiv},
       eprint = {2206.05864},
 primaryClass = {astro-ph.GA},
       adsurl = {https://ui.adsabs.harvard.edu/abs/2023A&A...674A..26C}
}

@ARTICLE{gaia_dr3,
       author = {{Gaia Collaboration} and {Vallenari}, A. and {Brown}, A.~G.~A. and {Prusti}, T. and {de Bruijne}, J.~H.~J. and {Arenou}, F. and {Babusiaux}, C. and {Biermann}, M. and {Creevey}, O.~L. and {Ducourant}, C. and {Evans}, D.~W. and {Eyer}, L. and {Guerra}, R. and {Hutton}, A. and {Jordi}, C. and {Klioner}, S.~A. and {Lammers}, U.~L. and {Lindegren}, L. and {Luri}, X. and {Mignard}, F. and {Panem}, C. and {Pourbaix}, D. and {Randich}, S. and {Sartoretti}, P. and {Soubiran}, C. and {Tanga}, P. and {Walton}, N.~A. and {Bailer-Jones}, C.~A.~L. and {Bastian}, U. and {Drimmel}, R. and {Jansen}, F. and {Katz}, D. and {Lattanzi}, M.~G. and {van Leeuwen}, F. and {Bakker}, J. and {Cacciari}, C. and {Casta{\~n}eda}, J. and {De Angeli}, F. and {Fabricius}, C. and {Fouesneau}, M. and {Fr{\'e}mat}, Y. and {Galluccio}, L. and {Guerrier}, A. and {Heiter}, U. and {Masana}, E. and {Messineo}, R. and {Mowlavi}, N. and {Nicolas}, C. and {Nienartowicz}, K. and {Pailler}, F. and {Panuzzo}, P. and {Riclet}, F. and {Roux}, W. and {Seabroke}, G.~M. and {Sordo}, R. and {Th{\'e}venin}, F. and {Gracia-Abril}, G. and {Portell}, J. and {Teyssier}, D. and {Altmann}, M. and {Andrae}, R. and {Audard}, M. and {Bellas-Velidis}, I. and {Benson}, K. and {Berthier}, J. and {Blomme}, R. and {Burgess}, P.~W. and {Busonero}, D. and {Busso}, G. and {C{\'a}novas}, H. and {Carry}, B. and {Cellino}, A. and {Cheek}, N. and {Clementini}, G. and {Damerdji}, Y. and {Davidson}, M. and {de Teodoro}, P. and {Nu{\~n}ez Campos}, M. and {Delchambre}, L. and {Dell'Oro}, A. and {Esquej}, P. and {Fern{\'a}ndez-Hern{\'a}ndez}, J. and {Fraile}, E. and {Garabato}, D. and {Garc{\'\i}a-Lario}, P. and {Gosset}, E. and {Haigron}, R. and {Halbwachs}, J. -L. and {Hambly}, N.~C. and {Harrison}, D.~L. and {Hern{\'a}ndez}, J. and {Hestroffer}, D. and {Hodgkin}, S.~T. and {Holl}, B. and {Jan{\ss}en}, K. and {Jevardat de Fombelle}, G. and {Jordan}, S. and {Krone-Martins}, A. and {Lanzafame}, A.~C. and {L{\"o}ffler}, W. and {Marchal}, O. and {Marrese}, P.~M. and {Moitinho}, A. and {Muinonen}, K. and {Osborne}, P. and {Pancino}, E. and {Pauwels}, T. and {Recio-Blanco}, A. and {Reyl{\'e}}, C. and {Riello}, M. and {Rimoldini}, L. and {Roegiers}, T. and {Rybizki}, J. and {Sarro}, L.~M. and {Siopis}, C. and {Smith}, M. and {Sozzetti}, A. and {Utrilla}, E. and {van Leeuwen}, M. and {Abbas}, U. and {{\'A}brah{\'a}m}, P. and {Abreu Aramburu}, A. and {Aerts}, C. and {Aguado}, J.~J. and {Ajaj}, M. and {Aldea-Montero}, F. and {Altavilla}, G. and {{\'A}lvarez}, M.~A. and {Alves}, J. and {Anders}, F. and {Anderson}, R.~I. and {Anglada Varela}, E. and {Antoja}, T. and {Baines}, D. and {Baker}, S.~G. and {Balaguer-N{\'u}{\~n}ez}, L. and {Balbinot}, E. and {Balog}, Z. and {Barache}, C. and {Barbato}, D. and {Barros}, M. and {Barstow}, M.~A. and {Bartolom{\'e}}, S. and {Bassilana}, J. -L. and {Bauchet}, N. and {Becciani}, U. and {Bellazzini}, M. and {Berihuete}, A. and {Bernet}, M. and {Bertone}, S. and {Bianchi}, L. and {Binnenfeld}, A. and {Blanco-Cuaresma}, S. and {Blazere}, A. and {Boch}, T. and {Bombrun}, A. and {Bossini}, D. and {Bouquillon}, S. and {Bragaglia}, A. and {Bramante}, L. and {Breedt}, E. and {Bressan}, A. and {Brouillet}, N. and {Brugaletta}, E. and {Bucciarelli}, B. and {Burlacu}, A. and {Butkevich}, A.~G. and {Buzzi}, R. and {Caffau}, E. and {Cancelliere}, R. and {Cantat-Gaudin}, T. and {Carballo}, R. and {Carlucci}, T. and {Carnerero}, M.~I. and {Carrasco}, J.~M. and {Casamiquela}, L. and {Castellani}, M. and {Castro-Ginard}, A. and {Chaoul}, L. and {Charlot}, P. and {Chemin}, L. and {Chiaramida}, V. and {Chiavassa}, A. and {Chornay}, N. and {Comoretto}, G. and {Contursi}, G. and {Cooper}, W.~J. and {Cornez}, T. and {Cowell}, S. and {Crifo}, F. and {Cropper}, M. and {Crosta}, M. and {Crowley}, C. and {Dafonte}, C. and {Dapergolas}, A. and {David}, M. and {David}, P. and {de Laverny}, P. and {De Luise}, F. and {De March}, R.},
        title = "{Gaia Data Release 3. Summary of the content and survey properties}",
      journal = {\aap},
         year = 2023,
        month = jun,
       volume = {674},
          eid = {A1},
        pages = {A1},
          doi = {10.1051/0004-6361/202243940},
archivePrefix = {arXiv},
       eprint = {2208.00211},
 primaryClass = {astro-ph.GA},
       adsurl = {https://ui.adsabs.harvard.edu/abs/2023A&A...674A...1G}
}

@ARTICLE{2025A&A...697A.177S,
       author = {{Santos}, A.~R.~G. and {Godoy-Rivera}, D. and {Mathur}, S. and {Breton}, S.~N. and {Garc{\'\i}a}, R.~A. and {Cunha}, M.~S.},
        title = "{Signature of spin-down stalling in stellar magnetic activity: The case of the open cluster NGC 6811}",
      journal = {\aap},
         year = 2025,
        month = may,
       volume = {697},
          eid = {A177},
        pages = {A177},
          doi = {10.1051/0004-6361/202554030},
archivePrefix = {arXiv},
       eprint = {2504.05413},
 primaryClass = {astro-ph.SR},
       adsurl = {https://ui.adsabs.harvard.edu/abs/2025A&A...697A.177S}
}

@ARTICLE{2023ApJ...951L..49C,
       author = {{Cao}, Lyra and {Pinsonneault}, Marc H. and {van Saders}, Jennifer L.},
        title = "{Core-envelope Decoupling Drives Radial Shear Dynamos in Cool Stars}",
      journal = {\apjl},
         year = 2023,
        month = jul,
       volume = {951},
       number = {2},
          eid = {L49},
        pages = {L49},
          doi = {10.3847/2041-8213/acd780},
archivePrefix = {arXiv},
       eprint = {2301.07716},
 primaryClass = {astro-ph.SR},
       adsurl = {https://ui.adsabs.harvard.edu/abs/2023ApJ...951L..49C}
}

@ARTICLE{Seli2025A&A...694A.161S,
       author = {{Seli}, B. and {Vida}, K. and {Ol{\'a}h}, K. and {G{\"o}rgei}, A. and {So{\'o}s}, Sz. and {P{\'a}l}, A. and {Kriskovics}, L. and {K{\H{o}}v{\'a}ri}, Zs.},
        title = "{Stellar flare morphology with TESS across the main sequence}",
      journal = {\aap},
         year = 2025,
        month = feb,
       volume = {694},
          eid = {A161},
        pages = {A161},
          doi = {10.1051/0004-6361/202452489},
archivePrefix = {arXiv},
       eprint = {2412.12989},
 primaryClass = {astro-ph.SR},
       adsurl = {https://ui.adsabs.harvard.edu/abs/2025A&A...694A.161S}
}

@ARTICLE{1998A&A...331...81P,
       author = {{Perryman}, M.~A.~C. and {Brown}, A.~G.~A. and {Lebreton}, Y. and {Gomez}, A. and {Turon}, C. and {Cayrel de Strobel}, G. and {Mermilliod}, J.~C. and {Robichon}, N. and {Kovalevsky}, J. and {Crifo}, F.},
        title = "{The Hyades: distance, structure, dynamics, and age}",
      journal = {\aap},
         year = 1998,
        month = mar,
       volume = {331},
        pages = {81-120},
          doi = {10.48550/arXiv.astro-ph/9707253},
archivePrefix = {arXiv},
       eprint = {astro-ph/9707253},
 primaryClass = {astro-ph},
       adsurl = {https://ui.adsabs.harvard.edu/abs/1998A&A...331...81P}
}

@ARTICLE{2009A&A...497..209V,
       author = {{van Leeuwen}, F.},
        title = "{Parallaxes and proper motions for 20 open clusters as based on the new Hipparcos catalogue}",
      journal = {\aap},
         year = 2009,
        month = apr,
       volume = {497},
       number = {1},
        pages = {209-242},
          doi = {10.1051/0004-6361/200811382},
archivePrefix = {arXiv},
       eprint = {0902.1039},
 primaryClass = {astro-ph.GA},
       adsurl = {https://ui.adsabs.harvard.edu/abs/2009A&A...497..209V}
}

@ARTICLE{2023A&A...677A.163A,
       author = {{Alfonso}, Jeison and {Garc{\'\i}a-Varela}, Alejandro},
        title = "{A Gaia astrometric view of the open clusters Pleiades, Praesepe, and Blanco 1}",
      journal = {\aap},
         year = 2023,
        month = sep,
       volume = {677},
          eid = {A163},
        pages = {A163},
          doi = {10.1051/0004-6361/202346569},
archivePrefix = {arXiv},
       eprint = {2304.00164},
 primaryClass = {astro-ph.GA},
       adsurl = {https://ui.adsabs.harvard.edu/abs/2023A&A...677A.163A}
}

@ARTICLE{2011AJ....141..172M,
       author = {{McArthur}, Barbara E. and {Benedict}, G. Fritz and {Harrison}, Thomas E. and {van Altena}, William},
        title = "{Astrometry with the Hubble Space Telescope: Trigonometric Parallaxes of Selected Hyads}",
      journal = {\aj},
         year = 2011,
        month = may,
       volume = {141},
       number = {5},
          eid = {172},
        pages = {172},
          doi = {10.1088/0004-6256/141/5/172},
archivePrefix = {arXiv},
       eprint = {1103.2094},
 primaryClass = {astro-ph.SR},
       adsurl = {https://ui.adsabs.harvard.edu/abs/2011AJ....141..172M}
}

@ARTICLE{2020MNRAS.493.4570L,
       author = {{Leitzinger}, M. and {Odert}, P. and {Greimel}, R. and {Vida}, K. and {Kriskovics}, L. and {Guenther}, E.~W. and {Korhonen}, H. and {Koller}, F. and {Hanslmeier}, A. and {K{\H{o}}v{\'a}ri}, Zs and {Lammer}, H.},
        title = "{A census of coronal mass ejections on solar-like stars}",
      journal = {\mnras},
         year = 2020,
        month = apr,
       volume = {493},
       number = {3},
        pages = {4570-4589},
          doi = {10.1093/mnras/staa504},
archivePrefix = {arXiv},
       eprint = {2002.04430},
 primaryClass = {astro-ph.SR},
       adsurl = {https://ui.adsabs.harvard.edu/abs/2020MNRAS.493.4570L}
}

@ARTICLE{2025ApJ...991...54C,
       author = {{Cao}, Dongtao and {Gu}, Shenghong},
        title = "{Doppler-shifted Signatures in the Chromospheric Line Profiles of II Pegasi}",
      journal = {\apj},
         year = 2025,
        month = sep,
       volume = {991},
       number = {1},
          eid = {54},
        pages = {54},
          doi = {10.3847/1538-4357/adf8eb},
       adsurl = {https://ui.adsabs.harvard.edu/abs/2025ApJ...991...54C}
}

@ARTICLE{2025A&A...694A.258R,
       author = {{Risbud}, Dhanraj and {Jadhav}, Vikrant V. and {Kroupa}, Pavel},
        title = "{Tidal tails of nearby open clusters: I. Mapping with Gaia DR3}",
      journal = {\aap},
         year = 2025,
        month = feb,
       volume = {694},
          eid = {A258},
        pages = {A258},
          doi = {10.1051/0004-6361/202453302},
archivePrefix = {arXiv},
       eprint = {2501.17225},
 primaryClass = {astro-ph.GA},
       adsurl = {https://ui.adsabs.harvard.edu/abs/2025A&A...694A.258R}
}

@ARTICLE{2024Univ...10..313V,
       author = {{Vida}, Kriszti{\'a}n and {K{\H{o}}v{\'a}ri}, {\mbox Zs}olt and {Leitzinger}, Martin and {Odert}, Petra and {Ol{\'a}h}, Katalin and {Seli}, B{\'a}lint and {Kriskovics}, Levente and {Greimel}, Robert and {G{\"o}rgei}, Anna M{\'a}ria},
        title = "{Stellar Flares, Superflares, and Coronal Mass Ejections{\textemdash}Entering the Big Data Era}",
      journal = {Universe},
         year = 2024,
        month = jul,
       volume = {10},
       number = {8},
          eid = {313},
        pages = {313},
          doi = {10.3390/universe10080313},
archivePrefix = {arXiv},
       eprint = {2407.16446},
 primaryClass = {astro-ph.SR},
       adsurl = {https://ui.adsabs.harvard.edu/abs/2024Univ...10..313V}
}

@ARTICLE{h_alpha2003,
       author = {{K{\"u}rster}, M. and {Endl}, M. and {Rouesnel}, F. and {Els}, S. and {Kaufer}, A. and {Brillant}, S. and {Hatzes}, A.~P. and {Saar}, S.~H. and {Cochran}, W.~D.},
        title = "{The low-level radial velocity variability   in Barnard's star (= GJ 699).  Secular acceleration, indications for convective redshift,  and planet mass limits}",
      journal = {\aap},
         year = 2003,
        month = jun,
       volume = {403},
        pages = {1077-1087},
          doi = {10.1051/0004-6361:20030396},
archivePrefix = {arXiv},
       eprint = {astro-ph/0303528},
 primaryClass = {astro-ph},
       adsurl = {https://ui.adsabs.harvard.edu/abs/2003A&A...403.1077K}
}

@ARTICLE{Namekataetal2022a,
       author = {{Namekata}, Kosuke and {Maehara}, Hiroyuki and {Honda}, Satoshi and {Notsu}, Yuta and {Okamoto}, Soshi and {Takahashi}, Jun and {Takayama}, Masaki and {Ohshima}, Tomohito and {Saito}, Tomoki and {Katoh}, Noriyuki and {Tozuka}, Miyako and {Murata}, Katsuhiro L. and {Ogawa}, Futa and {Niwano}, Masafumi and {Adachi}, Ryo and {Oeda}, Motoki and {Shiraishi}, Kazuki and {Isogai}, Keisuke and {Nogami}, Daisaku and {Shibata}, Kazunari},
        title = "{Discovery of a Long-duration Superflare on a Young Solar-type Star EK Draconis with Nearly Similar Time Evolution for H{\ensuremath{\alpha}} and White-light Emissions}",
      journal = {\apjl},
         year = 2022,
        month = feb,
       volume = {926},
       number = {1},
          eid = {L5},
        pages = {L5},
          doi = {10.3847/2041-8213/ac4df0},
archivePrefix = {arXiv},
       eprint = {2201.09416},
 primaryClass = {astro-ph.SR},
       adsurl = {https://ui.adsabs.harvard.edu/abs/2022ApJ...926L...5N}
}

@ARTICLE{2019A&A...622A.133I,
       author = {{Ilin}, Ekaterina and {Schmidt}, Sarah J. and {Davenport}, James R.~A. and {Strassmeier}, Klaus G.},
        title = "{Flares in open clusters with K2 . I. M 45 (Pleiades), M 44 (Praesepe), and M 67}",
      journal = {\aap},
         year = 2019,
        month = feb,
       volume = {622},
          eid = {A133},
        pages = {A133},
          doi = {10.1051/0004-6361/201834400},
archivePrefix = {arXiv},
       eprint = {1812.06725},
 primaryClass = {astro-ph.SR},
       adsurl = {https://ui.adsabs.harvard.edu/abs/2019A&A...622A.133I}
}

@ARTICLE{2022ApJ...931...45N,
       author = {{N{\'u}{\~n}ez}, Alejandro and {Ag{\"u}eros}, Marcel A. and {Covey}, Kevin R. and {Douglas}, Stephanie T. and {Drake}, Jeremy J. and {Rampalli}, Rayna and {Bowsher}, Emily C. and {Cargile}, Phillip A. and {Kraus}, Adam L. and {Law}, Nicholas M.},
        title = "{The Factory and the Beehive. IV. A Comprehensive Study of the Rotation X-Ray Activity Relation in Praesepe and the Hyades}",
      journal = {\apj},
         year = 2022,
        month = may,
       volume = {931},
       number = {1},
          eid = {45},
        pages = {45},
          doi = {10.3847/1538-4357/ac6517},
archivePrefix = {arXiv},
       eprint = {2205.06461},
 primaryClass = {astro-ph.SR},
       adsurl = {https://ui.adsabs.harvard.edu/abs/2022ApJ...931...45N}
}

@ARTICLE{2026ApJ...999..252T,
       author = {{Tran}, Andrew and {Song}, Inseok},
        title = "{Stellar Flare Study of nearby Young Moving Group Members with TESS Data}",
      journal = {\apj},
         year = 2026,
        month = mar,
       volume = {999},
       number = {2},
          eid = {252},
        pages = {252},
          doi = {10.3847/1538-4357/ae44f2},
archivePrefix = {arXiv},
       eprint = {2602.20402},
 primaryClass = {astro-ph.SR},
       adsurl = {https://ui.adsabs.harvard.edu/abs/2026ApJ...999..252T}
}

@ARTICLE{2024AJ....168...60F,
       author = {{Feinstein}, Adina D. and {Seligman}, Darryl Z. and {France}, Kevin and {Gagn{\'e}}, Jonathan and {Kowalski}, Adam},
        title = "{Evolution of Flare Activity in GKM Stars Younger Than 300 Myr over Five Years of TESS Observations}",
      journal = {\aj},
         year = 2024,
        month = aug,
       volume = {168},
       number = {2},
          eid = {60},
        pages = {60},
          doi = {10.3847/1538-3881/ad4edf},
archivePrefix = {arXiv},
       eprint = {2405.00850},
 primaryClass = {astro-ph.SR},
       adsurl = {https://ui.adsabs.harvard.edu/abs/2024AJ....168...60F}
}

@ARTICLE{2019ApJS..241...29Y,
       author = {{Yang}, Huiqin and {Liu}, Jifeng},
        title = "{The Flare Catalog and the Flare Activity in the Kepler Mission}",
      journal = {\apjs},
         year = 2019,
        month = apr,
       volume = {241},
       number = {2},
          eid = {29},
        pages = {29},
          doi = {10.3847/1538-4365/ab0d28},
archivePrefix = {arXiv},
       eprint = {1903.01056},
 primaryClass = {astro-ph.SR},
       adsurl = {https://ui.adsabs.harvard.edu/abs/2019ApJS..241...29Y}
}

@ARTICLE{2026AJ....171..163S,
       author = {{Stafford}, Matthew S. and {Curtis}, Jason Lee and {Ag{\"u}eros}, Marcel A.},
        title = "{Measuring Rotation Periods in Crowded Star Clusters with TESS: A Proof of Concept with NGC 3532}",
      journal = {\aj},
         year = 2026,
        month = mar,
       volume = {171},
       number = {3},
          eid = {163},
        pages = {163},
          doi = {10.3847/1538-3881/ae3c0c},
archivePrefix = {arXiv},
       eprint = {2601.17116},
 primaryClass = {astro-ph.SR},
       adsurl = {https://ui.adsabs.harvard.edu/abs/2026AJ....171..163S}
}

@ARTICLE{2020MNRAS.494.3766O,
       author = {{Odert}, P. and {Leitzinger}, M. and {Guenther}, E.~W. and {Heinzel}, P.},
        title = "{Stellar coronal mass ejections - II. Constraints from spectroscopic observations}",
      journal = {\mnras},
         year = 2020,
        month = may,
       volume = {494},
       number = {3},
        pages = {3766-3783},
          doi = {10.1093/mnras/staa1021},
archivePrefix = {arXiv},
       eprint = {2004.04063},
 primaryClass = {astro-ph.SR},
       adsurl = {https://ui.adsabs.harvard.edu/abs/2020MNRAS.494.3766O}
}

@ARTICLE{2016A&A...585A.150N,
       author = {{Netopil}, M. and {Paunzen}, E. and {Heiter}, U. and {Soubiran}, C.},
        title = "{On the metallicity of open clusters. III. Homogenised sample}",
      journal = {\aap},
         year = 2016,
        month = jan,
       volume = {585},
          eid = {A150},
        pages = {A150},
          doi = {10.1051/0004-6361/201526370},
archivePrefix = {arXiv},
       eprint = {1511.08884},
 primaryClass = {astro-ph.SR},
       adsurl = {https://ui.adsabs.harvard.edu/abs/2016A&A...585A.150N}
}

@ARTICLE{2018A&A...618A..93C,
       author = {{Cantat-Gaudin}, T. and {Jordi}, C. and {Vallenari}, A. and {Bragaglia}, A. and {Balaguer-N{\'u}{\~n}ez}, L. and {Soubiran}, C. and {Bossini}, D. and {Moitinho}, A. and {Castro-Ginard}, A. and {Krone-Martins}, A. and {Casamiquela}, L. and {Sordo}, R. and {Carrera}, R.},
        title = "{A Gaia DR2 view of the open cluster population in the Milky Way}",
      journal = {\aap},
         year = 2018,
        month = oct,
       volume = {618},
          eid = {A93},
        pages = {A93},
          doi = {10.1051/0004-6361/201833476},
archivePrefix = {arXiv},
       eprint = {1805.08726},
 primaryClass = {astro-ph.GA},
       adsurl = {https://ui.adsabs.harvard.edu/abs/2018A&A...618A..93C}
}

@INPROCEEDINGS{2024csss.confE..63V,
       author = {{Vida}, Kriszti{\'a}n and {Seli}, B{\'a}lint and {Roettenbacher}, Rachael M. and {G{\"o}rgei}, Anna M{\'a}ria and {Kriskovics}, Levente and {K{\H{o}}v{\'a}ri}, Zsolt and {Ol{\'a}h}, Katalin},
        title = "{Searching for stellar CMEs in the Praesepe and Pleiades clusters}",
    booktitle = {Cool Stars 22},
         year = 2024,
       series = {Cambridge Workshop on Cool Stars, Stellar Systems, and the Sun},
        month = jun,
          eid = {63},
        pages = {63},
          doi = {10.5281/zenodo.13132414},
       adsurl = {https://ui.adsabs.harvard.edu/abs/2024csss.confE..63V}
}

@ARTICLE{2019AJ....158..138S,
       author = {{Stassun}, Keivan G. and {Oelkers}, Ryan J. and {Paegert}, Martin and {Torres}, Guillermo and {Pepper}, Joshua and {De Lee}, Nathan and {Collins}, Kevin and {Latham}, David W. and {Muirhead}, Philip S. and {Chittidi}, Jay and {Rojas-Ayala}, B{\'a}rbara and {Fleming}, Scott W. and {Rose}, Mark E. and {Tenenbaum}, Peter and {Ting}, Eric B. and {Kane}, Stephen R. and {Barclay}, Thomas and {Bean}, Jacob L. and {Brassuer}, C.~E. and {Charbonneau}, David and {Ge}, Jian and {Lissauer}, Jack J. and {Mann}, Andrew W. and {McLean}, Brian and {Mullally}, Susan and {Narita}, Norio and {Plavchan}, Peter and {Ricker}, George R. and {Sasselov}, Dimitar and {Seager}, S. and {Sharma}, Sanjib and {Shiao}, Bernie and {Sozzetti}, Alessandro and {Stello}, Dennis and {Vanderspek}, Roland and {Wallace}, Geoff and {Winn}, Joshua N.},
        title = "{The Revised TESS Input Catalog and Candidate Target List}",
      journal = {\aj},
         year = 2019,
        month = oct,
       volume = {158},
       number = {4},
          eid = {138},
        pages = {138},
          doi = {10.3847/1538-3881/ab3467},
archivePrefix = {arXiv},
       eprint = {1905.10694},
 primaryClass = {astro-ph.SR},
       adsurl = {https://ui.adsabs.harvard.edu/abs/2019AJ....158..138S}
}

@ARTICLE{2017LRSP...14....4B,
       author = {{Brun}, Allan Sacha and {Browning}, Matthew K.},
        title = "{Magnetism, dynamo action and the solar-stellar connection}",
      journal = {Living Reviews in Solar Physics},
         year = 2017,
        month = dec,
       volume = {14},
       number = {1},
          eid = {4},
        pages = {4},
          doi = {10.1007/s41116-017-0007-8},
       adsurl = {https://ui.adsabs.harvard.edu/abs/2017LRSP...14....4B}
}

@ARTICLE{2023SSRv..219...70I,
       author = {{I{\textcommabelow s}{\i}k}, Emre and {van Saders}, Jennifer L. and {Reiners}, Ansgar and {Metcalfe}, Travis S.},
        title = "{Scaling and Evolution of Stellar Magnetic Activity}",
      journal = {\ssr},
         year = 2023,
        month = dec,
       volume = {219},
       number = {8},
          eid = {70},
        pages = {70},
          doi = {10.1007/s11214-023-01016-3},
archivePrefix = {arXiv},
       eprint = {2310.09515},
 primaryClass = {astro-ph.SR},
       adsurl = {https://ui.adsabs.harvard.edu/abs/2023SSRv..219...70I}
}

@ARTICLE{1986ApJS...61..531S,
       author = {{Stauffer}, J.~R. and {Hartmann}, L.~W.},
        title = "{Chromospheric Activity, Kinematics, and Metallicities of Nearby M Dwarfs}",
      journal = {\apjs},
         year = 1986,
        month = jul,
       volume = {61},
        pages = {531},
          doi = {10.1086/191123},
       adsurl = {https://ui.adsabs.harvard.edu/abs/1986ApJS...61..531S}
}

@ARTICLE{1979ApJ...234..579C,
       author = {{Cram}, L.~E. and {Mullan}, D.~J.},
        title = "{Model chromospheres of flare stars.I. Balmer-line profiles.}",
      journal = {\apj},
         year = 1979,
        month = dec,
       volume = {234},
        pages = {579-587},
          doi = {10.1086/157532},
       adsurl = {https://ui.adsabs.harvard.edu/abs/1979ApJ...234..579C}
}

@ARTICLE{2020ApJ...904..140C,
       author = {{Curtis}, Jason Lee and {Ag{\"u}eros}, Marcel A. and {Matt}, Sean P. and {Covey}, Kevin R. and {Douglas}, Stephanie T. and {Angus}, Ruth and {Saar}, Steven H. and {Cody}, Ann Marie and {Vanderburg}, Andrew and {Law}, Nicholas M. and {Kraus}, Adam L. and {Latham}, David W. and {Baranec}, Christoph and {Riddle}, Reed and {Ziegler}, Carl and {Lund}, Mikkel N. and {Torres}, Guillermo and {Meibom}, S{\o}ren and {Aguirre}, Victor Silva and {Wright}, Jason T.},
        title = "{When Do Stalled Stars Resume Spinning Down? Advancing Gyrochronology with Ruprecht 147}",
      journal = {\apj},
         year = 2020,
        month = dec,
       volume = {904},
       number = {2},
          eid = {140},
        pages = {140},
          doi = {10.3847/1538-4357/abbf58},
archivePrefix = {arXiv},
       eprint = {2010.02272},
 primaryClass = {astro-ph.SR},
       adsurl = {https://ui.adsabs.harvard.edu/abs/2020ApJ...904..140C}
}

@ARTICLE{2020ApJ...891..128M,
       author = {{Mullan}, D.~J. and {Houdebine}, E.~R.},
        title = "{A Transition of Dynamo Modes in M Dwarfs: Narrowing Down the Spectral Range Where the Transition Occurs}",
      journal = {\apj},
         year = 2020,
        month = mar,
       volume = {891},
       number = {2},
          eid = {128},
        pages = {128},
          doi = {10.3847/1538-4357/ab6ffa},
archivePrefix = {arXiv},
       eprint = {2001.08721},
 primaryClass = {astro-ph.SR},
       adsurl = {https://ui.adsabs.harvard.edu/abs/2020ApJ...891..128M}
}

@ARTICLE{2018ApJ...861L..11J,
       author = {{Jao}, Wei-Chun and {Henry}, Todd J. and {Gies}, Douglas R. and {Hambly}, Nigel C.},
        title = "{A Gap in the Lower Main Sequence Revealed by Gaia Data Release 2}",
      journal = {\apjl},
         year = 2018,
        month = jul,
       volume = {861},
       number = {1},
          eid = {L11},
        pages = {L11},
          doi = {10.3847/2041-8213/aacdf6},
archivePrefix = {arXiv},
       eprint = {1806.07792},
 primaryClass = {astro-ph.SR},
       adsurl = {https://ui.adsabs.harvard.edu/abs/2018ApJ...861L..11J}
}

@ARTICLE{2003ApJ...586..464B,
       author = {{Barnes}, Sydney A.},
        title = "{On the Rotational Evolution of Solar- and Late-Type Stars, Its Magnetic Origins, and the Possibility of Stellar Gyrochronology}",
      journal = {\apj},
         year = 2003,
        month = mar,
       volume = {586},
       number = {1},
        pages = {464-479},
          doi = {10.1086/367639},
archivePrefix = {arXiv},
       eprint = {astro-ph/0303631},
 primaryClass = {astro-ph},
       adsurl = {https://ui.adsabs.harvard.edu/abs/2003ApJ...586..464B}
}

@ARTICLE{2013ApJ...764..170D,
       author = {{Drake}, Jeremy J. and {Cohen}, Ofer and {Yashiro}, Seiji and {Gopalswamy}, Nat},
        title = "{Implications of Mass and Energy Loss due to Coronal Mass Ejections on Magnetically Active Stars}",
      journal = {\apj},
         year = 2013,
        month = feb,
       volume = {764},
       number = {2},
          eid = {170},
        pages = {170},
          doi = {10.1088/0004-637X/764/2/170},
archivePrefix = {arXiv},
       eprint = {1302.1136},
 primaryClass = {astro-ph.SR},
       adsurl = {https://ui.adsabs.harvard.edu/abs/2013ApJ...764..170D}
}

@ARTICLE{2012ApJ...759...71E,
       author = {{Emslie}, A.~G. and {Dennis}, B.~R. and {Shih}, A.~Y. and {Chamberlin}, P.~C. and {Mewaldt}, R.~A. and {Moore}, C.~S. and {Share}, G.~H. and {Vourlidas}, A. and {Welsch}, B.~T.},
        title = "{Global Energetics of Thirty-eight Large Solar Eruptive Events}",
      journal = {\apj},
         year = 2012,
        month = nov,
       volume = {759},
       number = {1},
          eid = {71},
        pages = {71},
          doi = {10.1088/0004-637X/759/1/71},
archivePrefix = {arXiv},
       eprint = {1209.2654},
 primaryClass = {astro-ph.SR},
       adsurl = {https://ui.adsabs.harvard.edu/abs/2012ApJ...759...71E}
}

@ARTICLE{2023SSRv..219...35C,
       author = {{Charbonneau}, Paul and {Sokoloff}, Dmitry},
        title = "{Evolution of Solar and Stellar Dynamo Theory}",
      journal = {\ssr},
         year = 2023,
        month = aug,
       volume = {219},
       number = {5},
          eid = {35},
        pages = {35},
          doi = {10.1007/s11214-023-00980-0},
archivePrefix = {arXiv},
       eprint = {2305.16553},
 primaryClass = {astro-ph.SR},
       adsurl = {https://ui.adsabs.harvard.edu/abs/2023SSRv..219...35C}
}

@BOOK{1980mfmd.book.....K,
       author = {{Krause}, F. and {Raedler}, K.-H.},
        title = "{Mean-field magnetohydrodynamics and dynamo theory}",
         year = 1980,
       adsurl = {https://ui.adsabs.harvard.edu/abs/1980mfmd.book.....K}
}

@ARTICLE{2018MNRAS.480.1711M,
       author = {{MacDonald}, James and {Gizis}, John},
        title = "{An explanation for the gap in the Gaia HRD for M dwarfs}",
      journal = {\mnras},
         year = 2018,
        month = oct,
       volume = {480},
       number = {2},
        pages = {1711-1714},
          doi = {10.1093/mnras/sty1888},
archivePrefix = {arXiv},
       eprint = {1806.11454},
 primaryClass = {astro-ph.SR},
       adsurl = {https://ui.adsabs.harvard.edu/abs/2018MNRAS.480.1711M}
}

@ARTICLE{1972ApJ...171..565S,
       author = {{Skumanich}, A.},
        title = "{Time Scales for Ca II Emission Decay, Rotational Braking, and Lithium Depletion}",
      journal = {\apj},
         year = 1972,
        month = feb,
       volume = {171},
        pages = {565},
          doi = {10.1086/151310},
       adsurl = {https://ui.adsabs.harvard.edu/abs/1972ApJ...171..565S}
}

@ARTICLE{1988ApJ...333..236K,
       author = {{Kawaler}, Steven D.},
        title = "{Angular Momentum Loss in Low-Mass Stars}",
      journal = {\apj},
         year = 1988,
        month = oct,
       volume = {333},
        pages = {236},
          doi = {10.1086/166740},
       adsurl = {https://ui.adsabs.harvard.edu/abs/1988ApJ...333..236K}
}

@ARTICLE{2015ApJ...799L..23M,
       author = {{Matt}, Sean P. and {Brun}, A. Sacha and {Baraffe}, Isabelle and {Bouvier}, J{\'e}r{\^o}me and {Chabrier}, Gilles},
        title = "{The Mass-dependence of Angular Momentum Evolution in Sun-like Stars}",
      journal = {\apjl},
         year = 2015,
        month = jan,
       volume = {799},
       number = {2},
          eid = {L23},
        pages = {L23},
          doi = {10.1088/2041-8205/799/2/L23},
archivePrefix = {arXiv},
       eprint = {1412.4786},
 primaryClass = {astro-ph.SR},
       adsurl = {https://ui.adsabs.harvard.edu/abs/2015ApJ...799L..23M}
}

@ARTICLE{2021LRSP...18....3V,
       author = {{Vidotto}, Aline A.},
        title = "{The evolution of the solar wind}",
      journal = {Living Reviews in Solar Physics},
         year = 2021,
        month = dec,
       volume = {18},
       number = {1},
          eid = {3},
        pages = {3},
          doi = {10.1007/s41116-021-00029-w},
archivePrefix = {arXiv},
       eprint = {2103.15748},
 primaryClass = {astro-ph.SR},
       adsurl = {https://ui.adsabs.harvard.edu/abs/2021LRSP...18....3V}
}

@ARTICLE{2020IJAsB..19..136A,
       author = {{Airapetian}, V.~S. and {Barnes}, R. and {Cohen}, O. and {Collinson}, G.~A. and {Danchi}, W.~C. and {Dong}, C.~F. and {Del Genio}, A.~D. and {France}, K. and {Garcia-Sage}, K. and {Glocer}, A. and {Gopalswamy}, N. and {Grenfell}, J.~L. and {Gronoff}, G. and {G{\"u}del}, M. and {Herbst}, K. and {Henning}, W.~G. and {Jackman}, C.~H. and {Jin}, M. and {Johnstone}, C.~P. and {Kaltenegger}, L. and {Kay}, C.~D. and {Kobayashi}, K. and {Kuang}, W. and {Li}, G. and {Lynch}, B.~J. and {L{\"u}ftinger}, T. and {Luhmann}, J.~G. and {Maehara}, H. and {Mlynczak}, M.~G. and {Notsu}, Y. and {Osten}, R.~A. and {Ramirez}, R.~M. and {Rugheimer}, S. and {Scheucher}, M. and {Schlieder}, J.~E. and {Shibata}, K. and {Sousa-Silva}, C. and {Stamenkovi{\'c}}, V. and {Strangeway}, R.~J. and {Usmanov}, A.~V. and {Vergados}, P. and {Verkhoglyadova}, O.~P. and {Vidotto}, A.~A. and {Voytek}, M. and {Way}, M.~J. and {Zank}, G.~P. and {Yamashiki}, Y.},
        title = "{Impact of space weather on climate and habitability of terrestrial-type exoplanets}",
      journal = {International Journal of Astrobiology},
         year = 2020,
        month = apr,
       volume = {19},
       number = {2},
        pages = {136-194},
          doi = {10.1017/S1473550419000132},
archivePrefix = {arXiv},
       eprint = {1905.05093},
 primaryClass = {astro-ph.EP},
       adsurl = {https://ui.adsabs.harvard.edu/abs/2020IJAsB..19..136A}
}

@ARTICLE{2016A&A...594A..95A,
       author = {{Alvarado-G{\'o}mez}, J.~D. and {Hussain}, G.~A.~J. and {Cohen}, O. and {Drake}, J.~J. and {Garraffo}, C. and {Grunhut}, J. and {Gombosi}, T.~I.},
        title = "{Simulating the environment around planet-hosting stars. II. Stellar winds and inner astrospheres}",
      journal = {\aap},
         year = 2016,
        month = oct,
       volume = {594},
          eid = {A95},
        pages = {A95},
          doi = {10.1051/0004-6361/201628988},
archivePrefix = {arXiv},
       eprint = {1607.08405},
 primaryClass = {astro-ph.SR},
       adsurl = {https://ui.adsabs.harvard.edu/abs/2016A&A...594A..95A}
}

@ARTICLE{2004A&ARv..12...71G,
       author = {{G{\"u}del}, Manuel},
        title = "{X-ray astronomy of stellar coronae}",
      journal = {\aapr},
         year = 2004,
        month = sep,
       volume = {12},
       number = {2-3},
        pages = {71-237},
          doi = {10.1007/s00159-004-0023-2},
archivePrefix = {arXiv},
       eprint = {astro-ph/0406661},
 primaryClass = {astro-ph},
       adsurl = {https://ui.adsabs.harvard.edu/abs/2004A&ARv..12...71G}
}

@ARTICLE{2007ApJ...662.1254S,
       author = {{Scholz}, Alexander and {Coffey}, Jaime and {Brandeker}, Alexis and {Jayawardhana}, Ray},
        title = "{Rotation and Activity of Pre-Main-Sequence Stars}",
      journal = {\apj},
         year = 2007,
        month = jun,
       volume = {662},
       number = {2},
        pages = {1254-1267},
          doi = {10.1086/518361},
archivePrefix = {arXiv},
       eprint = {0704.3266},
 primaryClass = {astro-ph},
       adsurl = {https://ui.adsabs.harvard.edu/abs/2007ApJ...662.1254S}
}

@ARTICLE{2010A&A...520A..15M,
       author = {{Messina}, S. and {Desidera}, S. and {Turatto}, M. and {Lanzafame}, A.~C. and {Guinan}, E.~F.},
        title = "{RACE-OC project: Rotation and variability of young stellar associations within 100 pc}",
      journal = {\aap},
         year = 2010,
        month = sep,
       volume = {520},
          eid = {A15},
        pages = {A15},
          doi = {10.1051/0004-6361/200913644},
archivePrefix = {arXiv},
       eprint = {1004.1959},
 primaryClass = {astro-ph.SR},
       adsurl = {https://ui.adsabs.harvard.edu/abs/2010A&A...520A..15M}
}

@ARTICLE{2006A&A...460L..35S,
       author = {{Stelzer}, B. and {Schmitt}, J.~H.~M.~M. and {Micela}, G. and {Liefke}, C.},
        title = "{Simultaneous optical and X-ray observations of a giant flare on the ultracool dwarf LP 412-31}",
      journal = {\aap},
         year = 2006,
        month = dec,
       volume = {460},
       number = {2},
        pages = {L35-L38},
          doi = {10.1051/0004-6361:20066488},
archivePrefix = {arXiv},
       eprint = {astro-ph/0610582},
 primaryClass = {astro-ph},
       adsurl = {https://ui.adsabs.harvard.edu/abs/2006A&A...460L..35S}
}

@ARTICLE{2011ApJ...731...63G,
       author = {{Gupta}, A. and {Galeazzi}, M. and {Williams}, B.},
        title = "{Contemporaneous XMM-Newton Investigation of a Giant X-ray Flare and Quiescent State from a Cool M-Class Dwarf in the Local Cavity}",
      journal = {\apj},
         year = 2011,
        month = apr,
       volume = {731},
       number = {1},
          eid = {63},
        pages = {63},
          doi = {10.1088/0004-637X/731/1/63},
archivePrefix = {arXiv},
       eprint = {1102.3696},
 primaryClass = {astro-ph.SR},
       adsurl = {https://ui.adsabs.harvard.edu/abs/2011ApJ...731...63G}
}

@ARTICLE{2016ApJ...828L..22S,
       author = {{Schmidt}, Sarah J. and {Shappee}, Benjamin J. and {Gagn{\'e}}, Jonathan and {Stanek}, K.~Z. and {Prieto}, Jos{\'e} L. and {Holoien}, Thomas W.-S. and {Kochanek}, C.~S. and {Chomiuk}, Laura and {Dong}, Subo and {Seibert}, Mark and {Strader}, Jay},
        title = "{ASASSN-16ae: A Powerful White-light Flare on an Early-L Dwarf}",
      journal = {\apjl},
         year = 2016,
        month = sep,
       volume = {828},
       number = {2},
          eid = {L22},
        pages = {L22},
          doi = {10.3847/2041-8205/828/2/L22},
archivePrefix = {arXiv},
       eprint = {1605.04313},
 primaryClass = {astro-ph.SR},
       adsurl = {https://ui.adsabs.harvard.edu/abs/2016ApJ...828L..22S}
}

@ARTICLE{2023ApJ...954..142L,
       author = {{Li}, Hua-Li and {Wang}, Jing and {Xin}, Li-Ping and {Bai}, Jian-Ying and {Han}, Xu-Hui and {Cai}, Hong-Bo and {Huang}, Lei and {Lu}, Xiao-Meng and {Qiu}, Yu-Lei and {Wu}, Chao and {Li}, Guang-Wei and {Deng}, Jing-Song and {Xu}, Da-Wei and {Yang}, Yuan-Gui and {Wang}, Xiang-Gao and {Liang}, En-Wei and {Wei}, Jian-Yan},
        title = "{White-light Superflare and Long-term Activity of the Nearby M7-type Binary EI Cnc Observed with GWAC System}",
      journal = {\apj},
         year = 2023,
        month = sep,
       volume = {954},
       number = {2},
          eid = {142},
        pages = {142},
          doi = {10.3847/1538-4357/ace59b},
archivePrefix = {arXiv},
       eprint = {2307.14594},
 primaryClass = {astro-ph.SR},
       adsurl = {https://ui.adsabs.harvard.edu/abs/2023ApJ...954..142L}
}

@ARTICLE{2024A&A...687A.138I,
       author = {{Ilin}, E. and {Poppenh{\"a}ger}, K. and {Stelzer}, B. and {Dsouza}, D.},
        title = "{The corona of a fully convective star with a near-polar flare}",
      journal = {\aap},
         year = 2024,
        month = jul,
       volume = {687},
          eid = {A138},
        pages = {A138},
          doi = {10.1051/0004-6361/202449541},
archivePrefix = {arXiv},
       eprint = {2405.05580},
 primaryClass = {astro-ph.SR},
       adsurl = {https://ui.adsabs.harvard.edu/abs/2024A&A...687A.138I}
}

@ARTICLE{2006SoPh..234...41K,
       author = {{Klimchuk}, James A.},
        title = "{On Solving the Coronal Heating Problem}",
      journal = {\solphys},
         year = 2006,
        month = mar,
       volume = {234},
       number = {1},
        pages = {41-77},
          doi = {10.1007/s11207-006-0055-z},
archivePrefix = {arXiv},
       eprint = {astro-ph/0511841},
 primaryClass = {astro-ph},
       adsurl = {https://ui.adsabs.harvard.edu/abs/2006SoPh..234...41K}
}

@BOOK{2011soca.book.....A,
       author = {{Aschwanden}, Markus J.},
        title = "{Self-Organized Criticality in Astrophysics}",
         year = 2011,
       adsurl = {https://ui.adsabs.harvard.edu/abs/2011soca.book.....A}
}

@ARTICLE{2015ApJ...809...79O,
       author = {{Osten}, Rachel A. and {Wolk}, Scott J.},
        title = "{Connecting Flares and Transient Mass-loss Events in Magnetically Active Stars}",
      journal = {\apj},
         year = 2015,
        month = aug,
       volume = {809},
       number = {1},
          eid = {79},
        pages = {79},
          doi = {10.1088/0004-637X/809/1/79},
archivePrefix = {arXiv},
       eprint = {1506.04994},
 primaryClass = {astro-ph.SR},
       adsurl = {https://ui.adsabs.harvard.edu/abs/2015ApJ...809...79O}
}

@ARTICLE{2015RSPTA.37340259T,
       author = {{Testa}, Paola and {Saar}, Steven H. and {Drake}, Jeremy J.},
        title = "{Stellar activity and coronal heating: an overview of recent results}",
      journal = {Philosophical Transactions of the Royal Society of London Series A},
         year = 2015,
        month = apr,
       volume = {373},
       number = {2042},
        pages = {20140259-20140259},
          doi = {10.1098/rsta.2014.0259},
archivePrefix = {arXiv},
       eprint = {1502.07401},
 primaryClass = {astro-ph.SR},
       adsurl = {https://ui.adsabs.harvard.edu/abs/2015RSPTA.37340259T}
}

@ARTICLE{2014PASP..126..398H,
       author = {{Howell}, Steve B. and {Sobeck}, Charlie and {Haas}, Michael and {Still}, Martin and {Barclay}, Thomas and {Mullally}, Fergal and {Troeltzsch}, John and {Aigrain}, Suzanne and {Bryson}, Stephen T. and {Caldwell}, Doug and {Chaplin}, William J. and {Cochran}, William D. and {Huber}, Daniel and {Marcy}, Geoffrey W. and {Miglio}, Andrea and {Najita}, Joan R. and {Smith}, Marcie and {Twicken}, J.~D. and {Fortney}, Jonathan J.},
        title = "{The K2 Mission: Characterization and Early Results}",
      journal = {\pasp},
         year = 2014,
        month = apr,
       volume = {126},
       number = {938},
        pages = {398},
          doi = {10.1086/676406},
archivePrefix = {arXiv},
       eprint = {1402.5163},
 primaryClass = {astro-ph.IM},
       adsurl = {https://ui.adsabs.harvard.edu/abs/2014PASP..126..398H}
}

@ARTICLE{2010ApJ...713L..79K,
       author = {{Koch}, David G. and {Borucki}, William J. and {Basri}, Gibor and {Batalha}, Natalie M. and {Brown}, Timothy M. and {Caldwell}, Douglas and {Christensen-Dalsgaard}, J{\o}rgen and {Cochran}, William D. and {DeVore}, Edna and {Dunham}, Edward W. and {Gautier}, III, Thomas N. and {Geary}, John C. and {Gilliland}, Ronald L. and {Gould}, Alan and {Jenkins}, Jon and {Kondo}, Yoji and {Latham}, David W. and {Lissauer}, Jack J. and {Marcy}, Geoffrey and {Monet}, David and {Sasselov}, Dimitar and {Boss}, Alan and {Brownlee}, Donald and {Caldwell}, John and {Dupree}, Andrea K. and {Howell}, Steve B. and {Kjeldsen}, Hans and {Meibom}, S{\o}ren and {Morrison}, David and {Owen}, Tobias and {Reitsema}, Harold and {Tarter}, Jill and {Bryson}, Stephen T. and {Dotson}, Jessie L. and {Gazis}, Paul and {Haas}, Michael R. and {Kolodziejczak}, Jeffrey and {Rowe}, Jason F. and {Van Cleve}, Jeffrey E. and {Allen}, Christopher and {Chandrasekaran}, Hema and {Clarke}, Bruce D. and {Li}, Jie and {Quintana}, Elisa V. and {Tenenbaum}, Peter and {Twicken}, Joseph D. and {Wu}, Hayley},
        title = "{Kepler Mission Design, Realized Photometric Performance, and Early Science}",
      journal = {\apjl},
         year = 2010,
        month = apr,
       volume = {713},
       number = {2},
        pages = {L79-L86},
          doi = {10.1088/2041-8205/713/2/L79},
archivePrefix = {arXiv},
       eprint = {1001.0268},
 primaryClass = {astro-ph.EP},
       adsurl = {https://ui.adsabs.harvard.edu/abs/2010ApJ...713L..79K}
}

@ARTICLE{2007JGRA..112.6112G,
       author = {{Gopalswamy}, N. and {Yashiro}, S. and {Akiyama}, S.},
        title = "{Geoeffectiveness of halo coronal mass ejections}",
      journal = {Journal of Geophysical Research (Space Physics)},
         year = 2007,
        month = jun,
       volume = {112},
       number = {A6},
          eid = {A06112},
        pages = {A06112},
          doi = {10.1029/2006JA012149},
       adsurl = {https://ui.adsabs.harvard.edu/abs/2007JGRA..112.6112G}
}

@ARTICLE{2012LRSP....9....3W,
       author = {{Webb}, David F. and {Howard}, Timothy A.},
        title = "{Coronal Mass Ejections: Observations}",
      journal = {Living Reviews in Solar Physics},
         year = 2012,
        month = dec,
       volume = {9},
       number = {1},
          eid = {3},
        pages = {3},
          doi = {10.12942/lrsp-2012-3},
       adsurl = {https://ui.adsabs.harvard.edu/abs/2012LRSP....9....3W}
}

@ARTICLE{2005JGRA..11012S05Y,
       author = {{Yashiro}, S. and {Gopalswamy}, N. and {Akiyama}, S. and {Michalek}, G. and {Howard}, R.~A.},
        title = "{Visibility of coronal mass ejections as a function of flare location and intensity}",
      journal = {Journal of Geophysical Research (Space Physics)},
         year = 2005,
        month = dec,
       volume = {110},
       number = {A12},
          eid = {A12S05},
        pages = {A12S05},
          doi = {10.1029/2005JA011151},
       adsurl = {https://ui.adsabs.harvard.edu/abs/2005JGRA..11012S05Y}
}

@ARTICLE{1994SoPh..154..275G,
       author = {{Garcia}, Howard A.},
        title = "{Temperature and Emission Measure from Goes Soft X-Ray Measurements}",
      journal = {\solphys},
         year = 1994,
        month = oct,
       volume = {154},
       number = {2},
        pages = {275-308},
          doi = {10.1007/BF00681100},
       adsurl = {https://ui.adsabs.harvard.edu/abs/1994SoPh..154..275G}
}

@INPROCEEDINGS{2011ASPC..448..841D,
       author = {{Delorme}, P. and {Cameron}, A.~C. and {Hebb}, L. and {Rostron}, J. and {Lister}, T.~A. and {Norton}, A.~J. and {Pollacco}, D. and {West}, R.~G.},
        title = "{Stellar Rotation in the Hyades and Praesepe: Gyrochronology and Braking Timescale}",
    booktitle = {16th Cambridge Workshop on Cool Stars, Stellar Systems, and the Sun},
         year = 2011,
       editor = {{Johns-Krull}, Christopher and {Browning}, Matthew K. and {West}, Andrew A.},
       series = {Astronomical Society of the Pacific Conference Series},
       volume = {448},
        month = dec,
        pages = {841},
       adsurl = {https://ui.adsabs.harvard.edu/abs/2011ASPC..448..841D}
}

@ARTICLE{2019ApJ...871..241D,
       author = {{Davenport}, James R.~A. and {Covey}, Kevin R. and {Clarke}, Riley W. and {Boeck}, Austin C. and {Cornet}, Jonathan and {Hawley}, Suzanne L.},
        title = "{The Evolution of Flare Activity with Stellar Age}",
      journal = {\apj},
         year = 2019,
        month = feb,
       volume = {871},
       number = {2},
          eid = {241},
        pages = {241},
          doi = {10.3847/1538-4357/aafb76},
archivePrefix = {arXiv},
       eprint = {1901.00890},
 primaryClass = {astro-ph.SR},
       adsurl = {https://ui.adsabs.harvard.edu/abs/2019ApJ...871..241D}
}

@ARTICLE{2007ApJ...666..393C,
       author = {{Cardini}, Daniela and {Cassatella}, Angelo},
        title = "{Color, Rotation, Age, and Chromospheric Activity Correlations in Late-Type Main-Sequence Stars}",
      journal = {\apj},
         year = 2007,
        month = sep,
       volume = {666},
       number = {1},
        pages = {393-402},
          doi = {10.1086/519991},
       adsurl = {https://ui.adsabs.harvard.edu/abs/2007ApJ...666..393C}
}

@ARTICLE{2025AJ....169..116L,
       author = {{Liu}, Rongrong and {Shao}, Zhengyi and {Li}, Lu},
        title = "{Photometric Determination of Unresolved Main-sequence Binaries in the Pleiades: Binary Fraction and Mass-ratio Distribution}",
      journal = {\aj},
         year = 2025,
        month = feb,
       volume = {169},
       number = {2},
          eid = {116},
        pages = {116},
          doi = {10.3847/1538-3881/ada380},
archivePrefix = {arXiv},
       eprint = {2501.01617},
 primaryClass = {astro-ph.SR},
       adsurl = {https://ui.adsabs.harvard.edu/abs/2025AJ....169..116L}
}

@ARTICLE{2016MNRAS.459.2408A,
       author = {{Aigrain}, S. and {Parviainen}, H. and {Pope}, B.~J.~S.},
        title = "{K2SC: flexible systematics correction and detrending of K2 light curves using Gaussian process regression}",
      journal = {\mnras},
         year = 2016,
        month = jul,
       volume = {459},
       number = {3},
        pages = {2408-2419},
          doi = {10.1093/mnras/stw706},
archivePrefix = {arXiv},
       eprint = {1603.09167},
 primaryClass = {astro-ph.SR},
       adsurl = {https://ui.adsabs.harvard.edu/abs/2016MNRAS.459.2408A}
}

@ARTICLE{2020RNAAS...4..204H,
       author = {{Huang}, Chelsea X. and {Vanderburg}, Andrew and {P{\'a}l}, Andras and {Sha}, Lizhou and {Yu}, Liang and {Fong}, Willie and {Fausnaugh}, Michael and {Shporer}, Avi and {Guerrero}, Natalia and {Vanderspek}, Roland and {Ricker}, George},
        title = "{Photometry of 10 Million Stars from the First Two Years of TESS Full Frame Images: Part I}",
      journal = {Research Notes of the American Astronomical Society},
         year = 2020,
        month = nov,
       volume = {4},
       number = {11},
          eid = {204},
        pages = {204},
          doi = {10.3847/2515-5172/abca2e},
archivePrefix = {arXiv},
       eprint = {2011.06459},
 primaryClass = {astro-ph.EP},
       adsurl = {https://ui.adsabs.harvard.edu/abs/2020RNAAS...4..204H}
}

@ARTICLE{2019A&A...623A..49V,
       author = {{Vida}, Kriszti{\'a}n and {Leitzinger}, Martin and {Kriskovics}, Levente and {Seli}, B{\'a}lint and {Odert}, Petra and {Kov{\'a}cs}, Orsolya Eszter and {Korhonen}, Heidi and {van Driel-Gesztelyi}, Lidia},
        title = "{The quest for stellar coronal mass ejections in late-type stars. I. Investigating Balmer-line asymmetries of single stars in Virtual Observatory data}",
      journal = {\aap},
         year = 2019,
        month = mar,
       volume = {623},
          eid = {A49},
        pages = {A49},
          doi = {10.1051/0004-6361/201834264},
archivePrefix = {arXiv},
       eprint = {1901.04229},
 primaryClass = {astro-ph.SR},
       adsurl = {https://ui.adsabs.harvard.edu/abs/2019A&A...623A..49V}
}

@ARTICLE{2021NatAs...5..697V,
       author = {{Veronig}, Astrid M. and {Odert}, Petra and {Leitzinger}, Martin and {Dissauer}, Karin and {Fleck}, Nikolaus C. and {Hudson}, Hugh S.},
        title = "{Indications of stellar coronal mass ejections through coronal dimmings}",
      journal = {Nature Astronomy},
         year = 2021,
        month = jan,
       volume = {5},
        pages = {697-706},
          doi = {10.1038/s41550-021-01345-9},
archivePrefix = {arXiv},
       eprint = {2110.12029},
 primaryClass = {astro-ph.SR},
       adsurl = {https://ui.adsabs.harvard.edu/abs/2021NatAs...5..697V}
}

@ARTICLE{2025Natur.647..603C,
       author = {{Callingham}, J.~R. and {Tasse}, C. and {Keers}, R. and {Kavanagh}, R.~D. and {Vedantham}, H.~K. and {Zarka}, P. and {Bellotti}, S. and {Cristofari}, P.~I. and {Bloot}, S. and {Konijn}, D.~C. and {Hardcastle}, M.~J. and {Lamy}, L. and {Pass}, E.~K. and {Pope}, B.~J.~S. and {Reid}, H. and {R{\"o}ttgering}, H.~J.~A. and {Shimwell}, T.~W. and {Zucca}, P.},
        title = "{Radio burst from a stellar coronal mass ejection}",
      journal = {\nat},
         year = 2025,
        month = nov,
       volume = {647},
       number = {8090},
        pages = {603-607},
          doi = {10.1038/s41586-025-09715-3},
archivePrefix = {arXiv},
       eprint = {2511.09289},
 primaryClass = {astro-ph.SR},
       adsurl = {https://ui.adsabs.harvard.edu/abs/2025Natur.647..603C}
}

@ARTICLE{2018A&A...616A.163V,
       author = {{Vida}, Kriszti{\'a}n and {Roettenbacher}, Rachael M.},
        title = "{Finding flares in Kepler data using machine-learning tools}",
      journal = {\aap},
         year = 2018,
        month = sep,
       volume = {616},
          eid = {A163},
        pages = {A163},
          doi = {10.1051/0004-6361/201833194},
archivePrefix = {arXiv},
       eprint = {1806.00334},
 primaryClass = {astro-ph.SR},
       adsurl = {https://ui.adsabs.harvard.edu/abs/2018A&A...616A.163V}
}

\appendix

\section{FFDs}

\begin{figure*}
    \centering
    \includegraphics[width=0.95\linewidth]{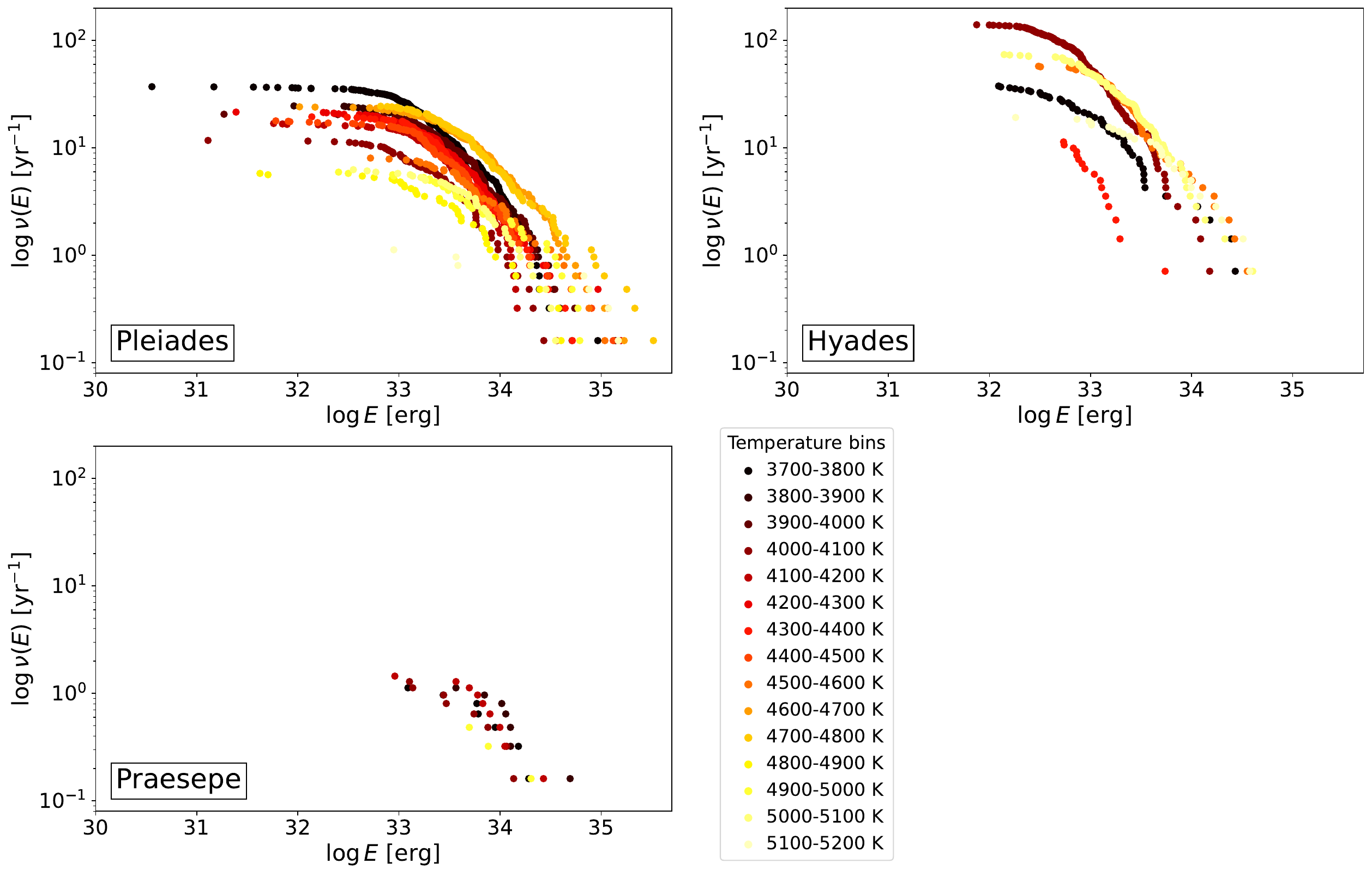}
    \caption{FFDs for 100 K temperature bins ranging from 3700 K to 5200 K. No pattern can be see in the change of the flare rate with effective temperature.}
    \label{fig:Kbins}
\end{figure*}

\section{Dynamic spectra}

\begin{figure*}
\centering
\begin{tabular}{cc}
\includegraphics[width=0.37\linewidth]{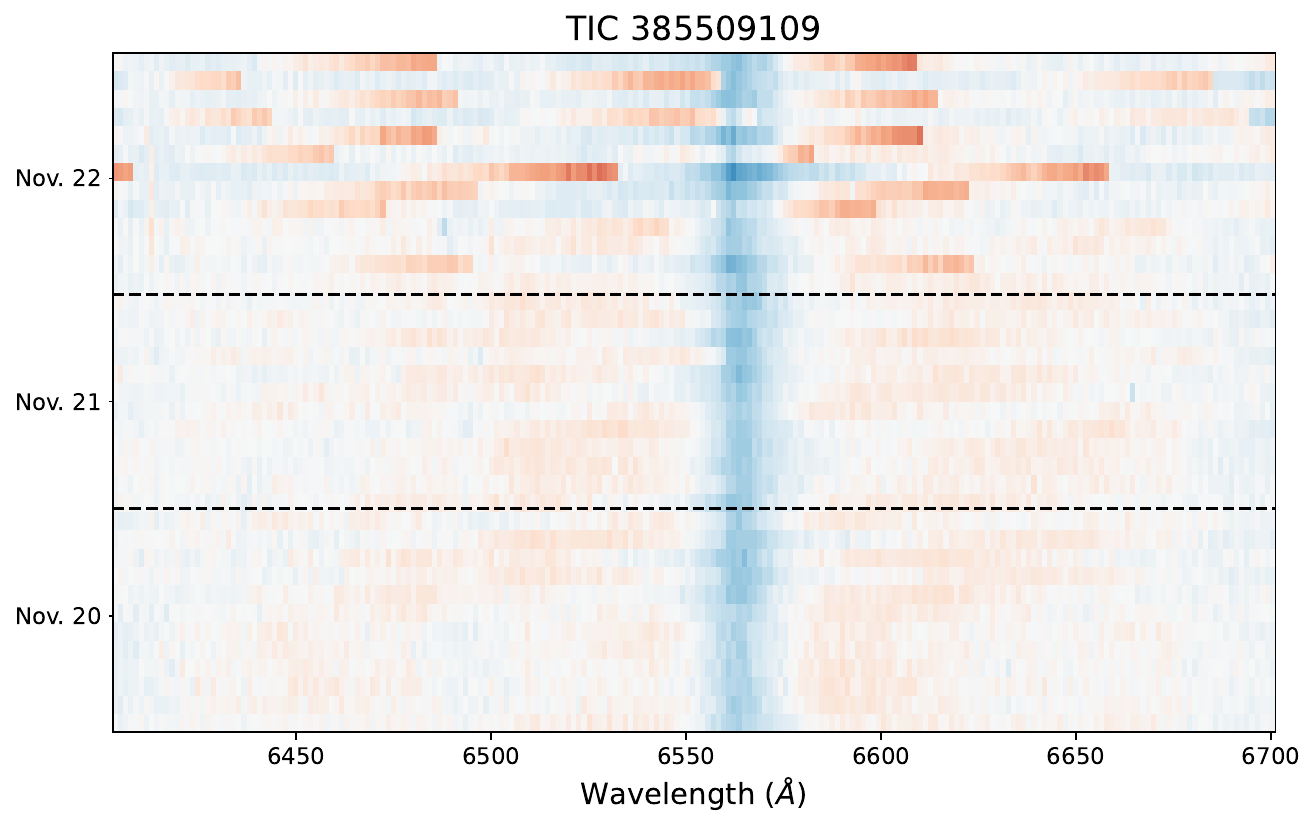}&
\includegraphics[width=0.37\linewidth]{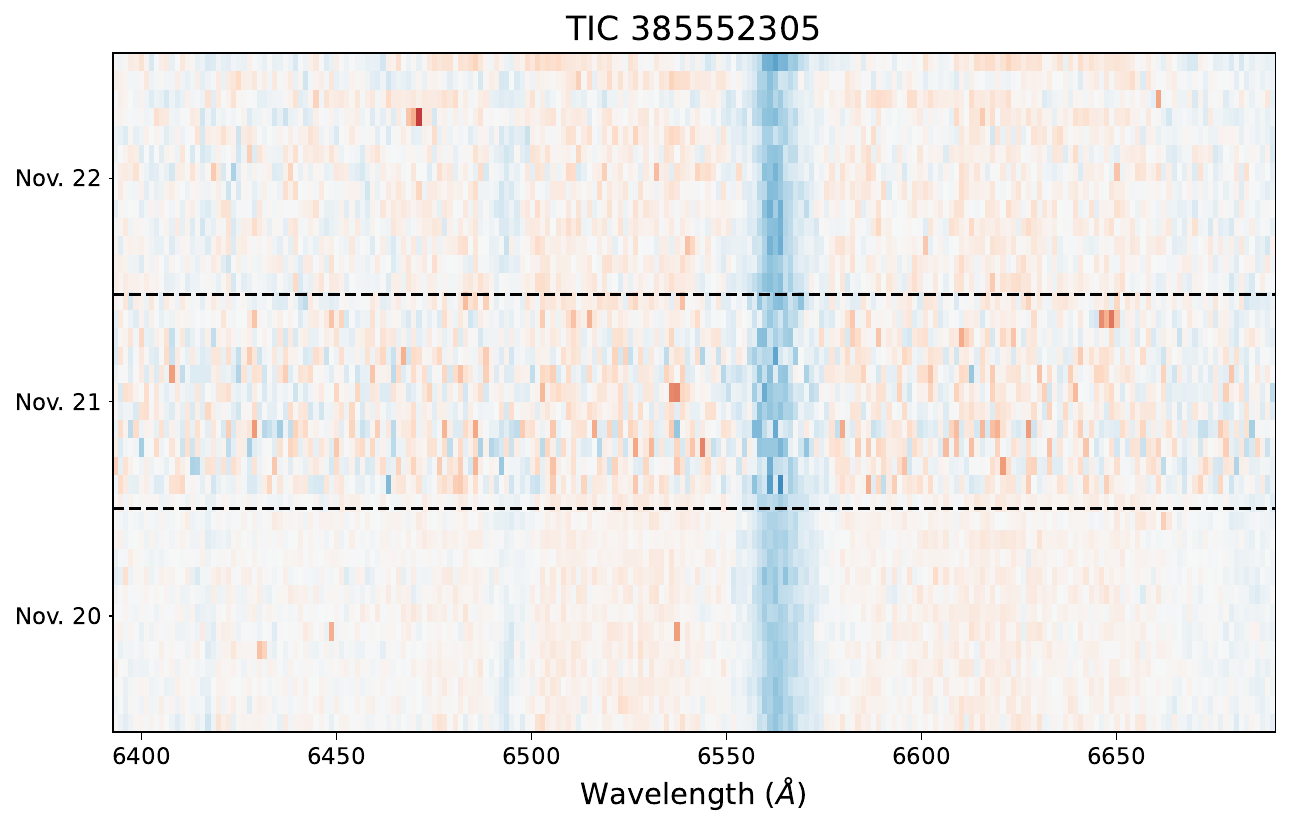}\\

\includegraphics[width=0.37\linewidth]{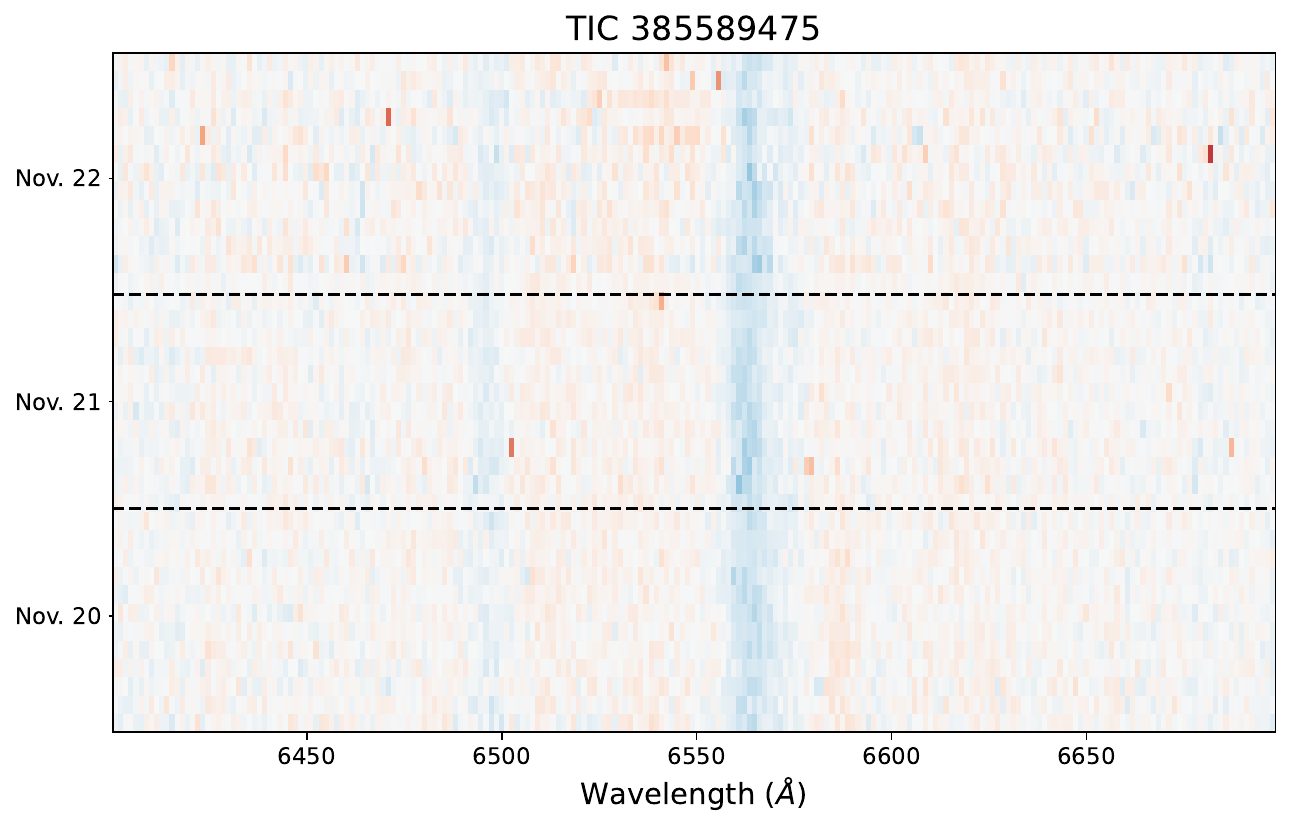}&
\includegraphics[width=0.37\linewidth]{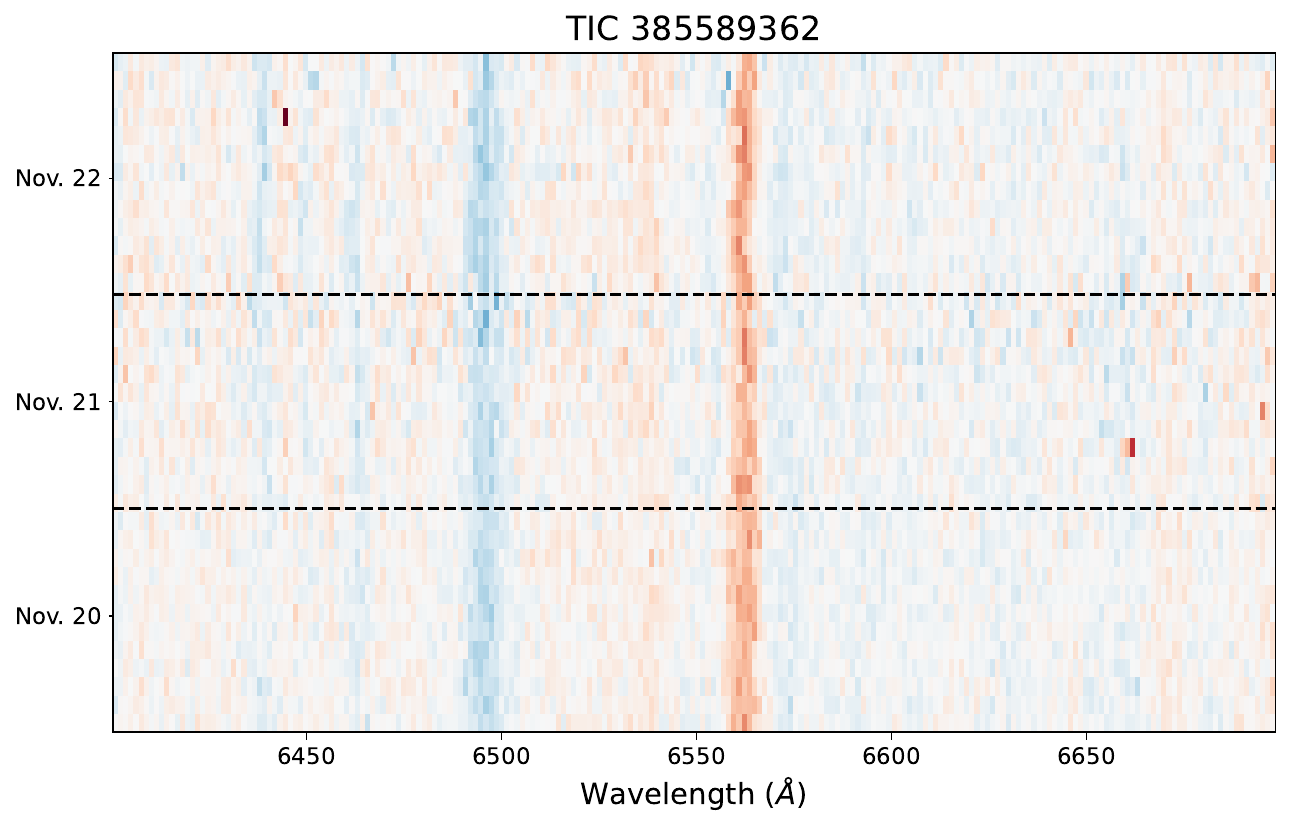}\\

\end{tabular}
\caption{Dynamic spectra of four active stars in the Pleiades cluster with MDM/OSMOS. The wavelength range is focused on the H$\alpha$ line. The color scale goes from blue (absorption) to red (emission). The spectra of TIC 385509109 shows signs of of emission at various wavelengths. These are caused by instrumental effects. No sign of flare or CME activity is visible.}
\label{fig:ple_spec}

\end{figure*}

\begin{figure*}
\centering
\begin{tabular}{cc}
\includegraphics[width=0.37\linewidth]{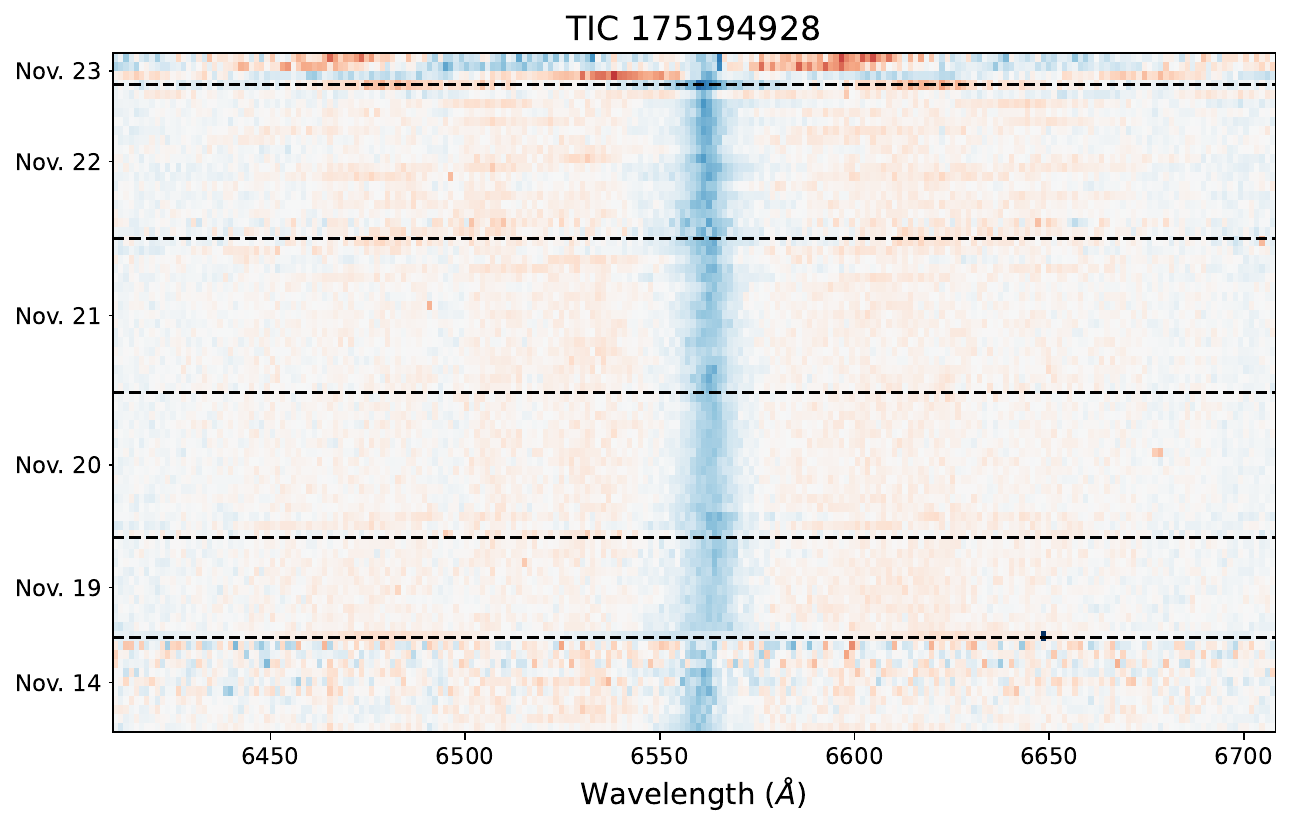}&\includegraphics[width=0.37\linewidth]{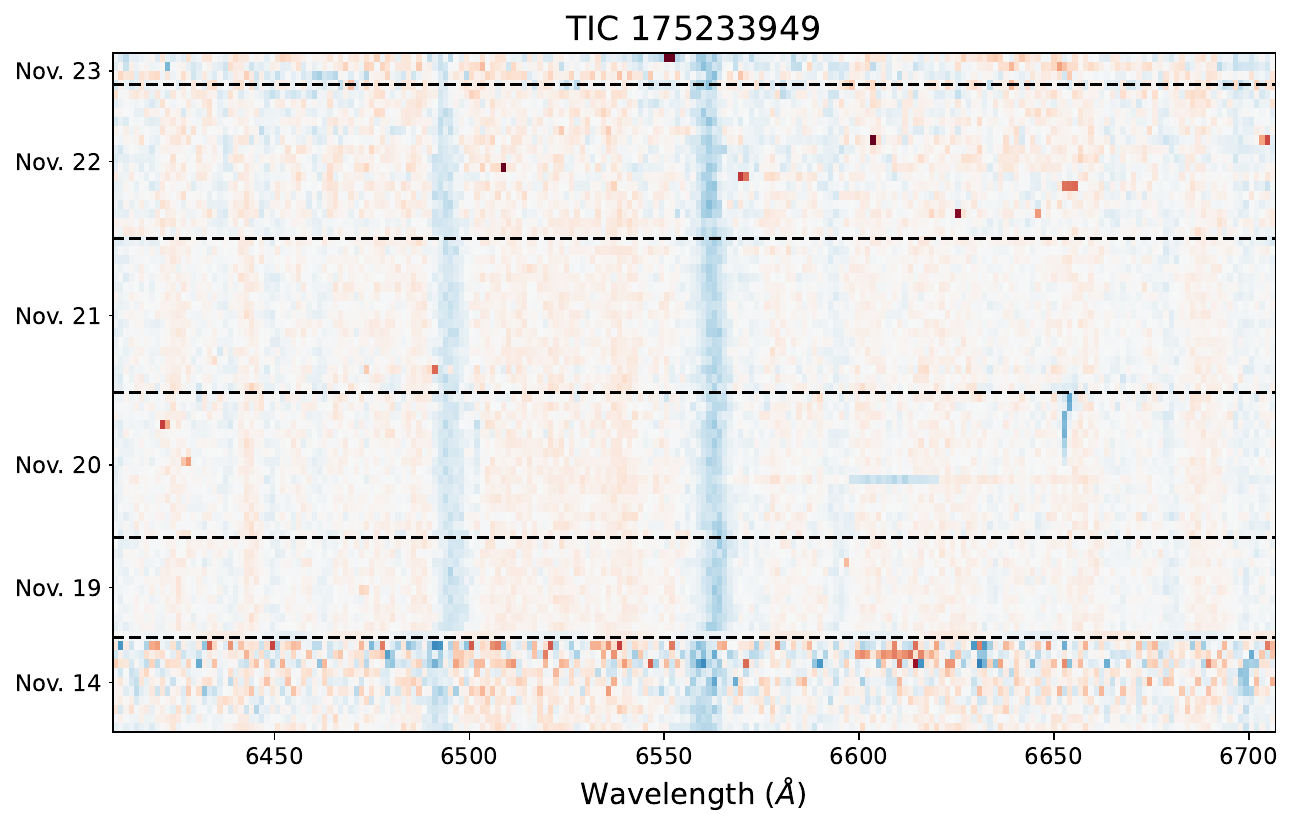}\\

\includegraphics[width=0.37\linewidth]{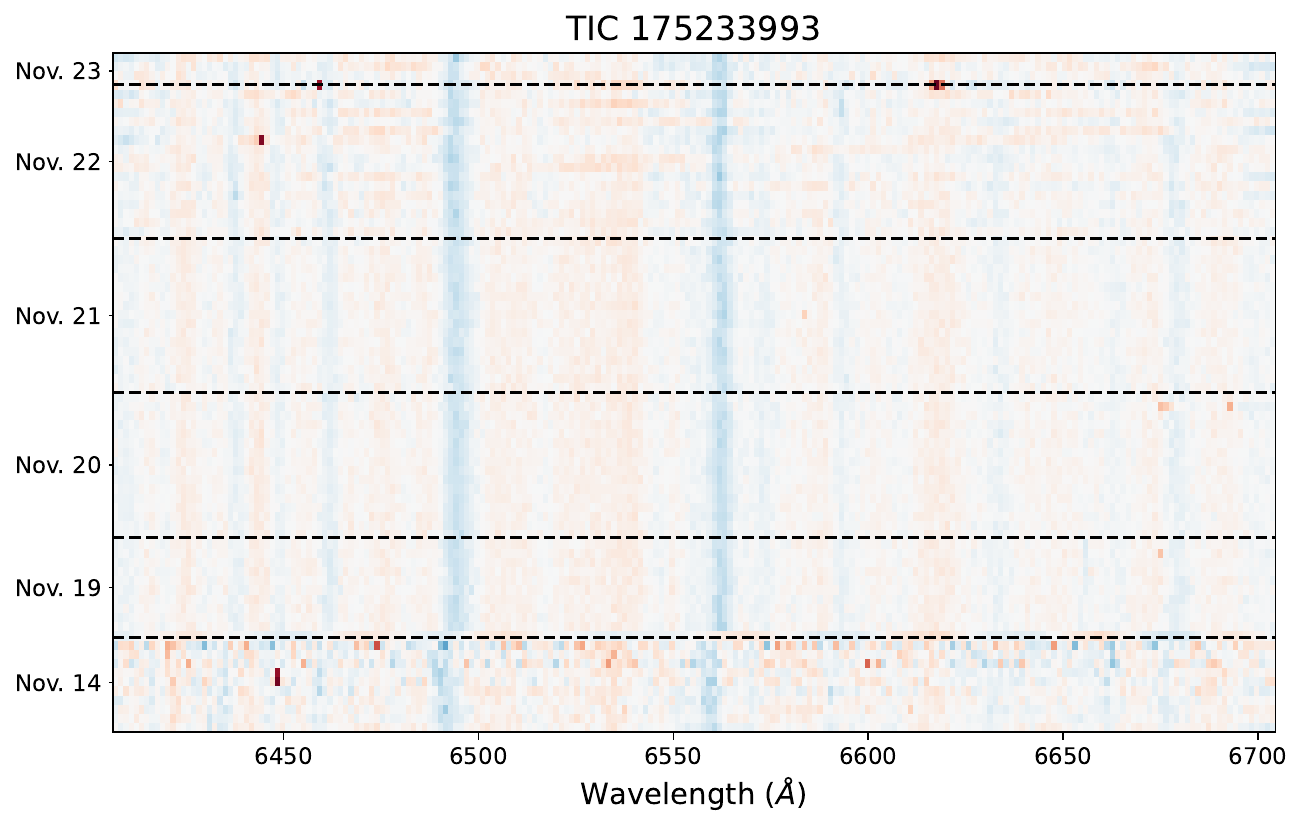}&\includegraphics[width=0.37\linewidth]{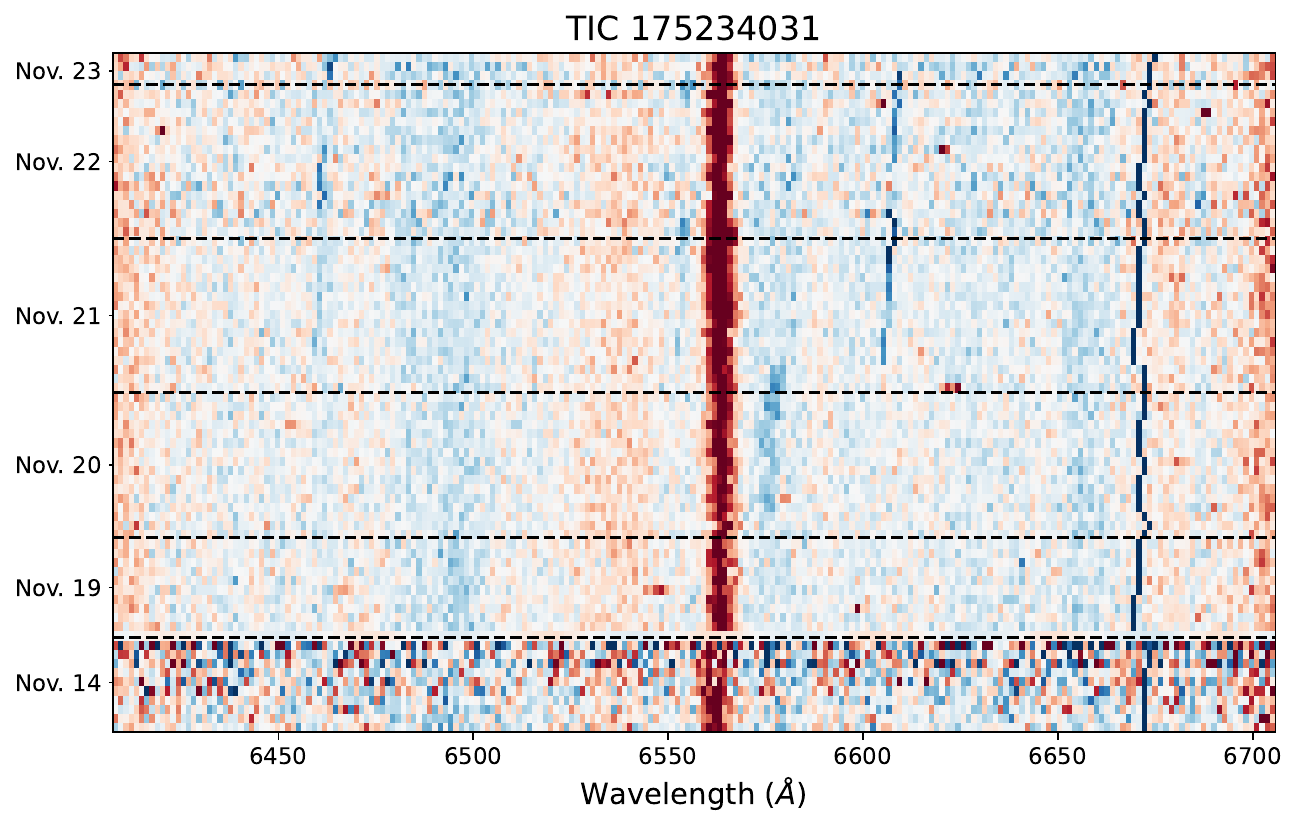}\\

\includegraphics[width=0.37\linewidth]{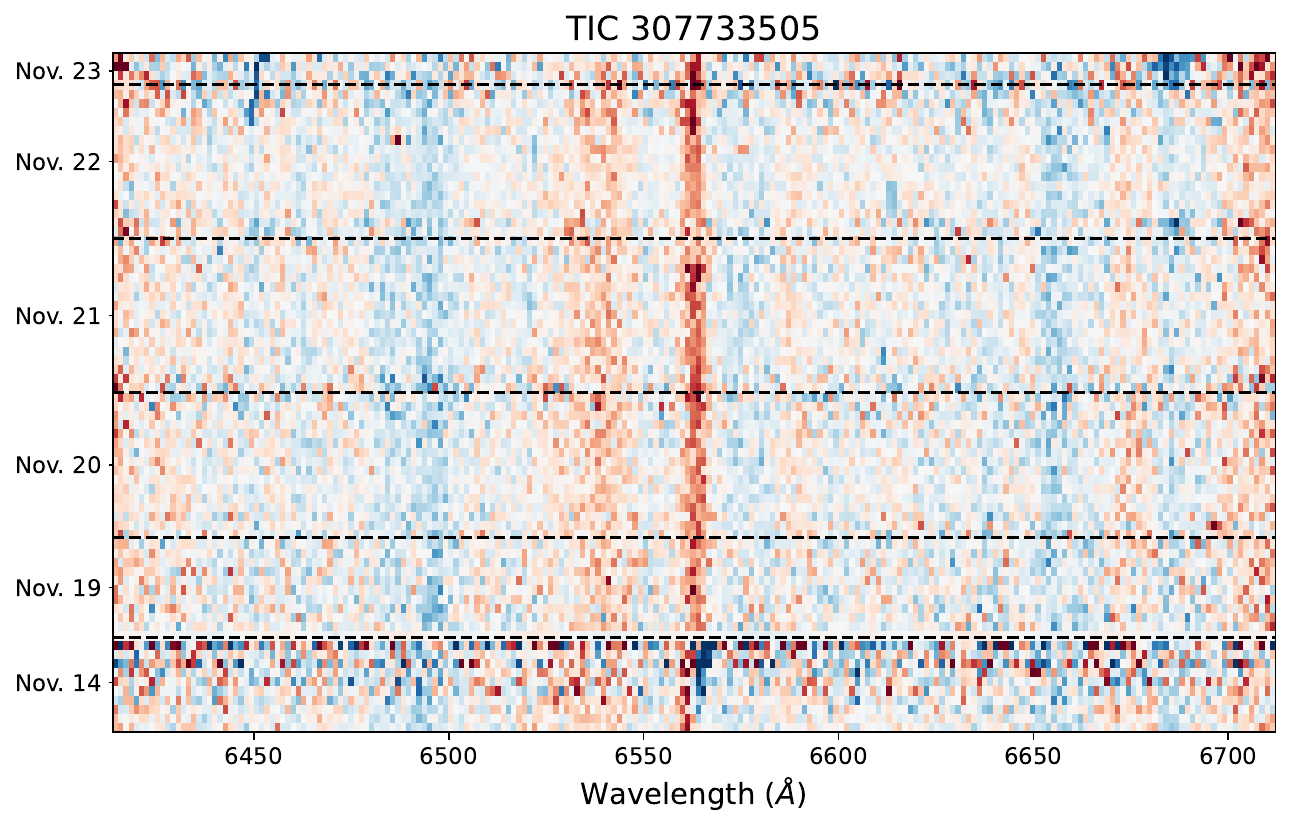}& \\

\end{tabular}
\caption{Same as Fig.~\ref{fig:ple_spec}, but for the five active stars in the Praesepe cluster. No signs of flare or CME activity are visible.}
\label{fig:pra_spec}

\end{figure*}

\section{Color--rotation diagrams}



\begin{figure}
    \centering
    \includegraphics[width=1\linewidth]{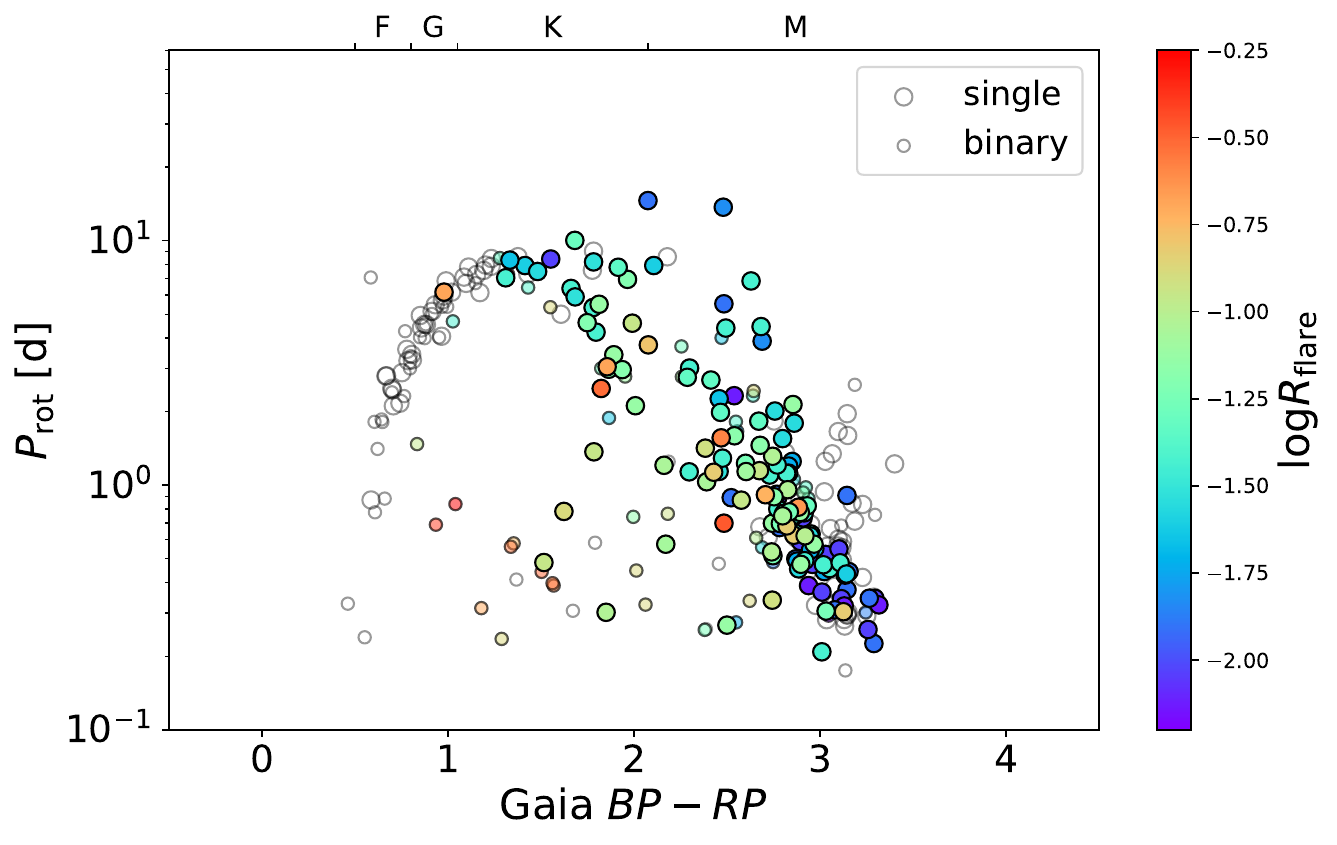}
    \includegraphics[width=1\linewidth]{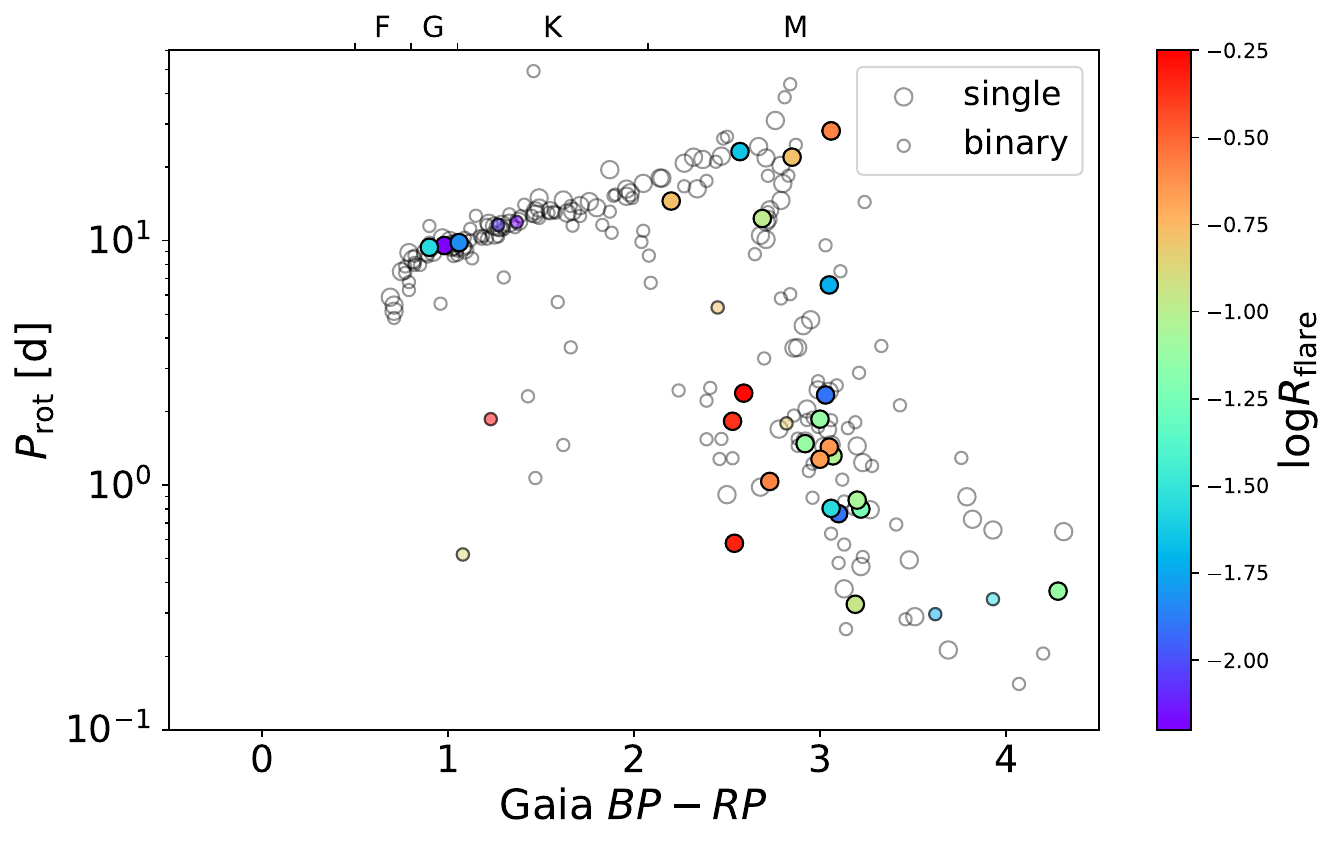}
    \includegraphics[width=1\linewidth]{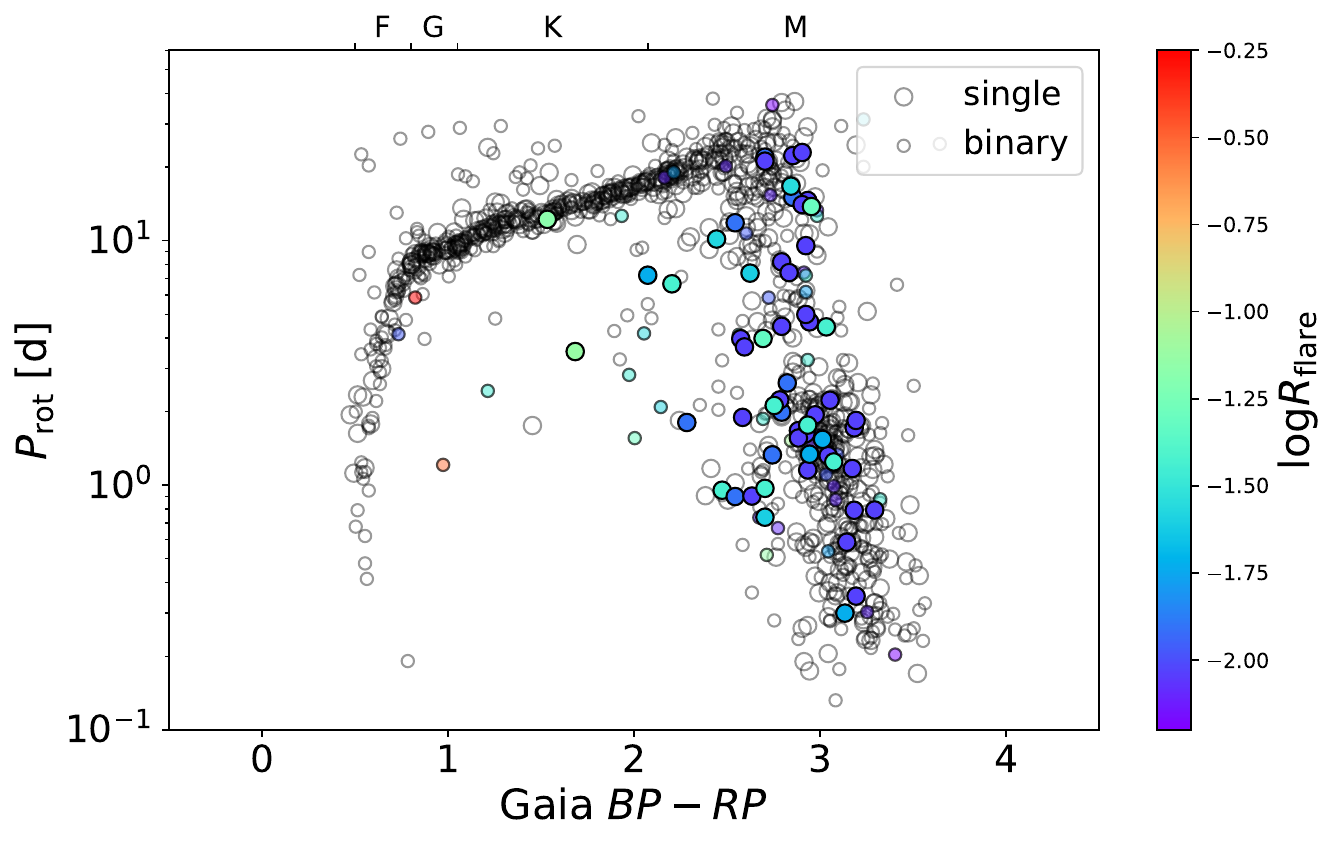}
    \caption{Color--rotation diagram of the three analyzed clusters with extinction corrected Gaia $BP-RP$ index. The top, middle and bottom panels show the Pleiades, the Hyades and Praesepe, respectively. All cluster members are marked by black outlined dots. Single and binary stars are differentiated by the size of the marker, dots corresponding to single stars have larger radius. The flaring stars are colored according to their yearly flare rates ($R_{\rm flare}$).}
    \label{fig:prot_vs_color3}
\end{figure}

\end{document}